\documentclass[12pt,preprint]{aastex}

\newcommand{\bet}{$\rm \beta~$}
\newcommand{\bets}{$\rm \beta$}

\def\gs{\mathrel{\raise0.35ex\hbox{$\scriptstyle >$}\kern-0.6em \lower0.40ex\hbox{{$\scriptstyle \sim$}}}}
\def\ls{\mathrel{\raise0.35ex\hbox{$\scriptstyle <$}\kern-0.6em \lower0.40ex\hbox{{$\scriptstyle \sim$}}}}

\newcommand{\Qv}{$\rm Q_{\nu}~$}

\newcommand{\um}{$\,\mu$m~}
\newcommand{\ums}{$\,\mu$m}

\newcommand{\m}{$\rm \,$m~}

\newcommand{\mm}{$\rm \,$mm~}

\newcommand{\K}{$\rm \,K~$}
\newcommand{\Ks}{$\rm \,K$}

\newcommand{\Lsun}{$\rm \,L_\odot~$}

\newcommand{\Msun}{$\rm \,M_\odot~$}
\newcommand{\Msuns}{$\rm \,M_\odot$}

\newcommand{\DL}{$\rm D_{\tiny{L}}~$}

\newcommand{\Td}{$\rm T_d~$}
\newcommand{\Tds}{$\rm T_d$}

\newcommand{\Lfir}{$\rm L_{\mbox{\tiny{FIR}}}~$}
\newcommand{\Lfirs}{$\rm L_{\mbox{\tiny{FIR}}}$}

\newcommand{\Md}{$\rm M_d~$}
\newcommand{\Mds}{$\rm M_d$}

\newcommand{\phrva}{PhRvA}

\begin{document}
\title{350\um Observations of Local Luminous Infrared Galaxies and the Temperature Dependence of the Emissivity Index}
\author{M. Yang and T. Phillips}
\affil{Caltech, Department of Physics, MC 320-47, Pasadena, CA 91125}
\email{min@submm.caltech.edu}

\begin{abstract}
We report 350\um observations of 18 nearby luminous infrared galaxies (LIRGs),
using the Submillimeter High Angular Resolution Camera II 
(SHARC-II) mounted on the Caltech Submillimeter Observatory (CSO) 10.4\m telescope. 
Combining our 350\um flux measurements with the existing far-infrared (far-IR)
and submillimeter (submm) data, we fit a single-temperature model to 
the spectral energy distributions (SEDs), and find the dust temperatures, emissivity indices 
and far-IR luminosities 
having sample medians of \mbox{$\rm T_d = 39.4 \pm 7.9 \,\rm K$}, \mbox{$\rm \beta = 1.6 \pm 0.3$} 
and \mbox{$\rm L_{\mbox{\tiny{FIR}}} = 10^{\,11.2 \pm 0.6} \,L_\odot$}.
An empirical inverse \mbox{\Tds-\bet} correlation, 
best described by \mbox{$\rm T_d  =  [9.86 \times 10^{9}] \, ^ {\frac{1}{4.63 + \beta}}$},
is established for the local LIRG sample, which we argue 
can be explained by the intrinsic interdependence 
between the dust temperature and grain emissivity index 
as physical parameters, as well as variations
in grain properties in the the interstellar medium (ISM).

\end{abstract}

\keywords{dust --- galaxies: infrared --- submillimeter: galaxies}

\section{Introduction}\label{intro}

Interstellar dust grains are small particles ($\sim 0.01 - 0.1 \rm \, \mu m$) that 
sparsely populate the ISM. 
Although interstellar dust accounts for only a very small 
fraction of the total mass in a galaxy, it plays critical 
roles in galaxy formation and evolution. Dust grains absorb 
strongly in the ultraviolet (UV) and optical, 
leading to a significant fraction of stellar radiation within 
a galaxy being absorbed. The warm grains subsequently emit 
strongly in the far-IR/submm, effectively down-converting the 
electromagnetic energy in various astrophysical environments.
Observations of dust emission at far-IR and submm wavelengths 
are crucial in probing physical conditions and star formation activity. 

Emission and absorption line features are generally weak in the 
far-IR/submm (Blain et al.\ 2003), leading to smooth far-IR/submm 
SEDs dominated by dust thermal continuum emission.
However, modeling of the far-IR/submm SEDs observed in astronomical 
objects is far from trivial, as the observed emission spectrum is a complex 
function of radiative transfer as well as distributions in grain properties, such as 
composition, size and shape, that affect the way dust absorbs and emits radiation.
In the simple case of a uniform grain population, dust thermal 
emission is well-approximated 
by a graybody (modified blackbody) function (Hildebrand\ 1983) 
\begin{equation}
\rm S_\nu = \Omega \, B_\nu(T_d) \, Q_\nu \label{sed}
\,,\end{equation}
where \Td is the dust temperature and \Qv is the absorption coefficient.
Eq.~(\ref{sed}) implicitly incorporates Kirchhoff's law, which ensures 
the equality between the emissivity and the absorption coefficient 
at all frequencies. In the far-IR/submm, 
$\rm Q_{\nu} = Q_0 \, (\frac{\nu}{\nu_{0}})^\beta$, where $\rm Q_{0}$ is the absorption
coefficient normalized at some reference frequency  $\nu_{0}$ and \bet is the emissivity index
\footnote{A blackbody has absorption coefficient $\rm Q_{\nu} = 1$ and emissivity index \bet = 0 
at all frequencies.}.
There are more elaborate far-IR/submm SED models 
that attempt to account for multiple dust components and general optical depths. 
However, the applications of such complex models are impractical 
when the frequency sampling of the SED is limited. More importantly, 
Blain et al.\ (2003) show that the added complexity, even if feasible, 
generally does not lead to appreciable differences in constraining the observed far-IR/submm SEDs. 
Note that the SED parameters estimated by adopting the single-temperature SED model
are \emph{effective}, brightness-weighted average values from
complex mixtures of dust grains with different temperatures, properties and
optical depths in the ISM.

Dust temperature profiles vary significantly for grain populations with 
distinctly different radii (i.e., a few submicrons versus tens of angstroms). 
The very small grains ($a \leq 50 \,\rm{\AA} $)
undergo large temperature fluctuations on absorption of a photon (Sellgren et al.\ 1985), while 
the larger ``standard'' grains reach equilibrium 
temperatures, as determined by thermal equilibrium between absorption of UV/optical photons 
and emission of far-IR/submm photons (Greenberg\ 1978), i.e., 
\begin{equation}
\int_{\rm{UV/optical}} \pi \,a^2 {\rm Q_\nu} \, c \, u_\nu \, d\nu   =  
4 \pi \int_{\rm{far-IR/submm}} {\rm B_\nu(T_d)} \, \pi a^2 \, {\rm Q_\nu} \, d\nu \label{equi_e1}\,,
\end{equation}
where 
$u_\nu$ is the energy density of the incident radiation field. 
To derive the exact solution to Eq.~(\ref{equi_e1}), one would need to have 
accurate knowledge about the actual \Qv values at all frequencies -
information that is lacking at present. At optical wavelengths, we know 
\Qv is relatively constant and near unity, and it is generally assumed that 
on average \Qv = 0.5 in the UV/optical in the numerical integration 
of the LHS of Eq.~(\ref{equi_e1}) (Martin\ 1978). While on the RHS of Eq.~(\ref{equi_e1}),
$\rm Q_\nu  =  Q_0 \,(\nu/\nu_0)^\beta$ (by the definition of the emissivity index \bets) 
at far-IR/submm wavelengths. For a uniform grain population, 
we thus have the approximate relation
\begin{equation}
{\rm T_d} \propto (\frac{\rm F}{\rm Q_{0}})^{\rm \frac{1}{4+\beta}}\label{equi_td1}\,,
\end{equation}
where $\rm F$ is the integrated 
incident flux, defined as ${\rm F} \equiv \int_{\rm{UV/optical}} c\, u_\nu \, d\nu $,
characterizing the strength of the incident radiation field
\footnote{
$\rm F = \frac{8\,\pi\,h\,\nu_0^4}{c^2}\,Q_0\,({k\,T_d}/{h\,\nu_0})^{4+\beta}\,\Gamma(4+\beta)\,\zeta(4+\beta)$, 
where \mbox{$\rm \Gamma(z) = \int_0^\infty t^{z-1}e^{-t}\,dt$} 
and \mbox{$\rm \zeta(s) = \frac{1}{\Gamma(s)} \int_0^\infty \frac{t^{s-1}}{e^t-1}\,dt$} 
are the Gamma and Riemann $\zeta$ functions respectively. This formula 
is equivalent to that given by De Breuck et al.\ (2003).}.

Accurate knowledge of the emissivity index, which characterizes 
the frequency dependence of the absorption coefficient of interstellar dust, 
is crucial for determining important large-scale characteristics of dusty galaxies, 
such as dust temperature, dust mass, and the existence of multiple dust components. 
Given that the emissivity index is a complex function of various grain properties, its estimate
also offers insights into grain properties, particularly grain structure (Seki \& Yamamoto\ 1980; Yang\ 2006)
and formation of ice mantles (Aannestad 1975). 
While theory predicts \bet values of 1 or 2 for interstellar dust in the long-wavelength limit,
observations of astronomical objects in the far-IR/submm
have revealed a wide range of \bet values. 
Lower emissivity indices have been observed in active 
galactic environments, such as circumstellar disks and warm molecular 
clouds, as well as large-scale dust distributions in infrared luminous 
dusty galaxies (Knapp et al.\ 1993; Blake et al.\ 1996; Dunne et al.\ 2000). 
Far-IR emissivity that has a frequency 
dependence steeper than quadratic has also been observed in some
galactic sources (Schwartz\ 1982; Lis et al.\ 1998). 
In particular, Dupac et al.\ (2003) reported a wide range in \bet values, 
as well as a negative temperature dependence of \bets,
as observed in a large sample of Galactic molecular clouds.

\section{Sample and Observations}\label{local}
Sources in this sample are mostly selected from the LIRG sample presented by 
Wynn-William \& Becklin (WWB, 1993). The WWB sample, 
not being a strictly defined complete sample, includes 19 sources 
that were selected from the {\it IRAS\,} Point Source Catalog
for their high luminosities and high flux densities at 12 and 25\ums. 
16 sources here are selected from the WWB sample by requiring 
$\rm S_{100\, \mu m} \geq 20\,Jy$. In addition, two ULIRGs - UGC\,5101 and Mrk\,231 - are added 
to the list to better understand nearby infrared galaxies at the high luminosity end. 
In total, our sample includes 18 nearby LIRGs in the redshift range of $0.003 \leq z \leq 0.042$.
This sample, with its selection based on high luminosity and high flux densities 
at the {\it IRAS\,} wavebands, is likely to have an enhanced fraction of galaxies 
that are more active and contain warmer dust as compared to the bulk of the submm
population at similar redshifts. A wealth of photometry data at far-IR/submm/mm wavelengths 
is available from the literature for a majority of the sources, 
which enables us to estimate both dust temperature and emissivity index 
in the SED fitting. Table\ \ref{local-target} and Table\ \ref{local-photometry} list the sample and the 
existing far-IR/submm/mm photometric data. 

\begin{figure}
\begin{center}
\begin{tabular}{cccc}
{\includegraphics[width=1.5in, angle=270]{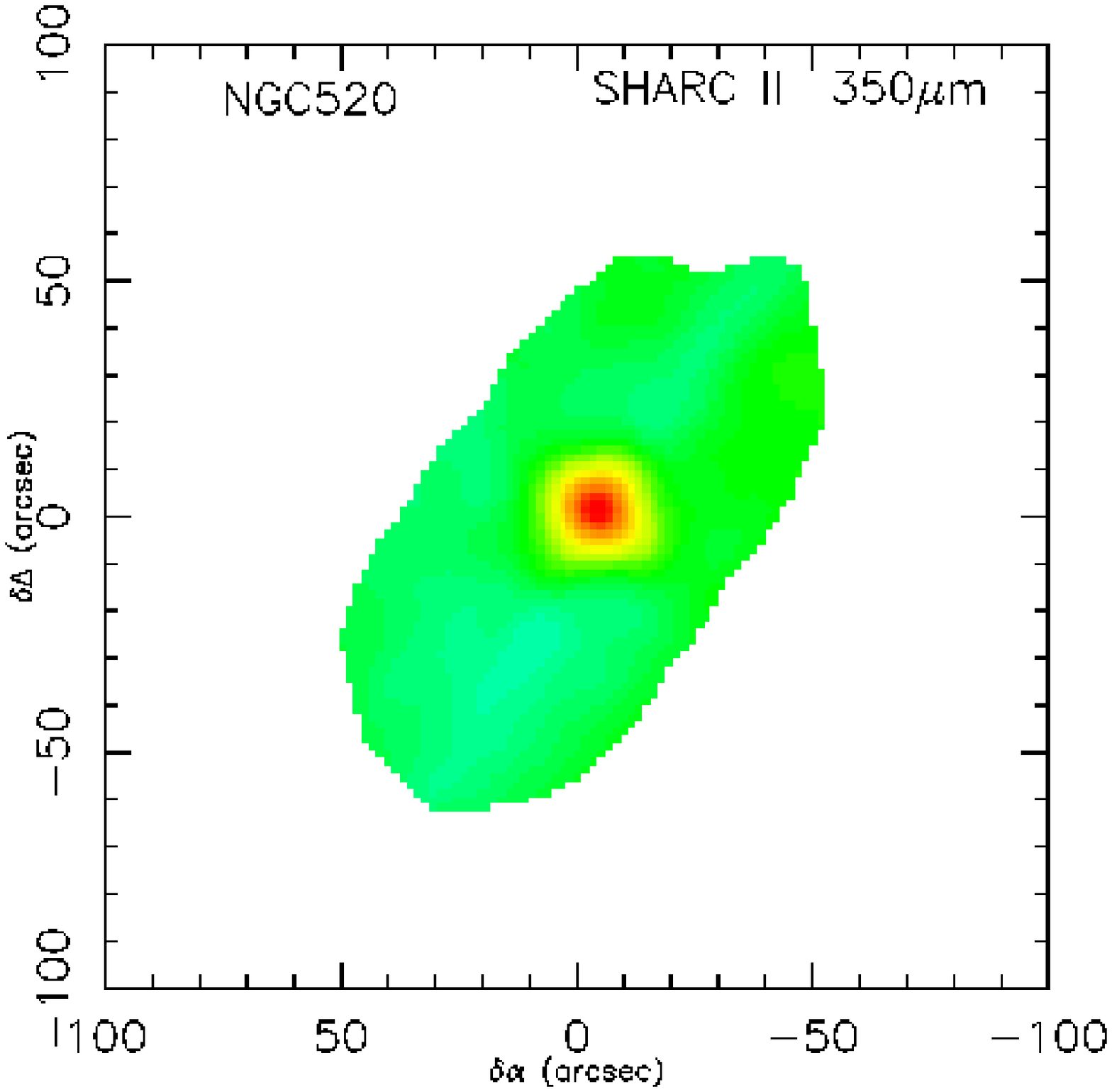}}&
{\includegraphics[width=1.5in, angle=270]{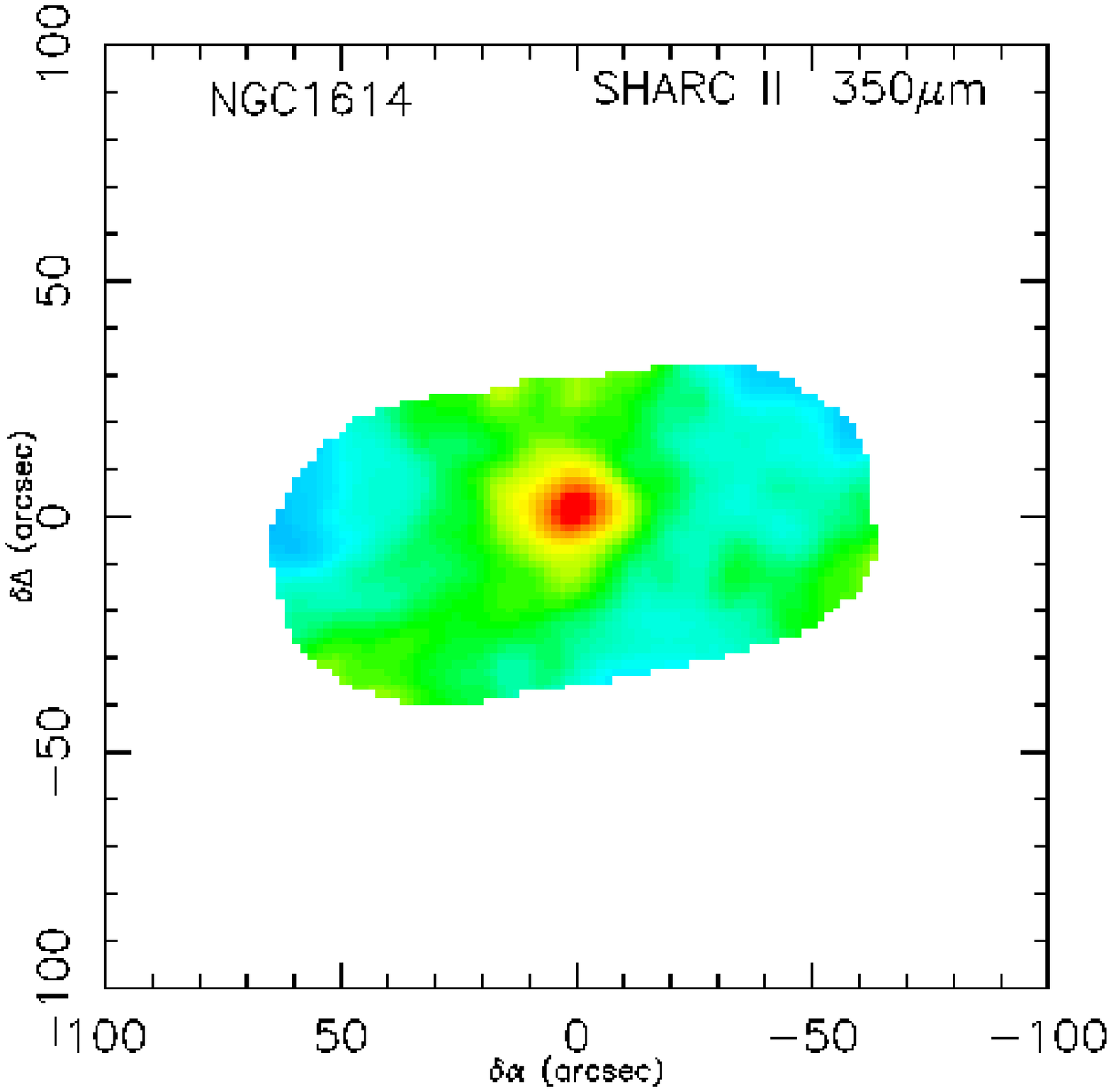}}&
{\includegraphics[width=1.5in, angle=270]{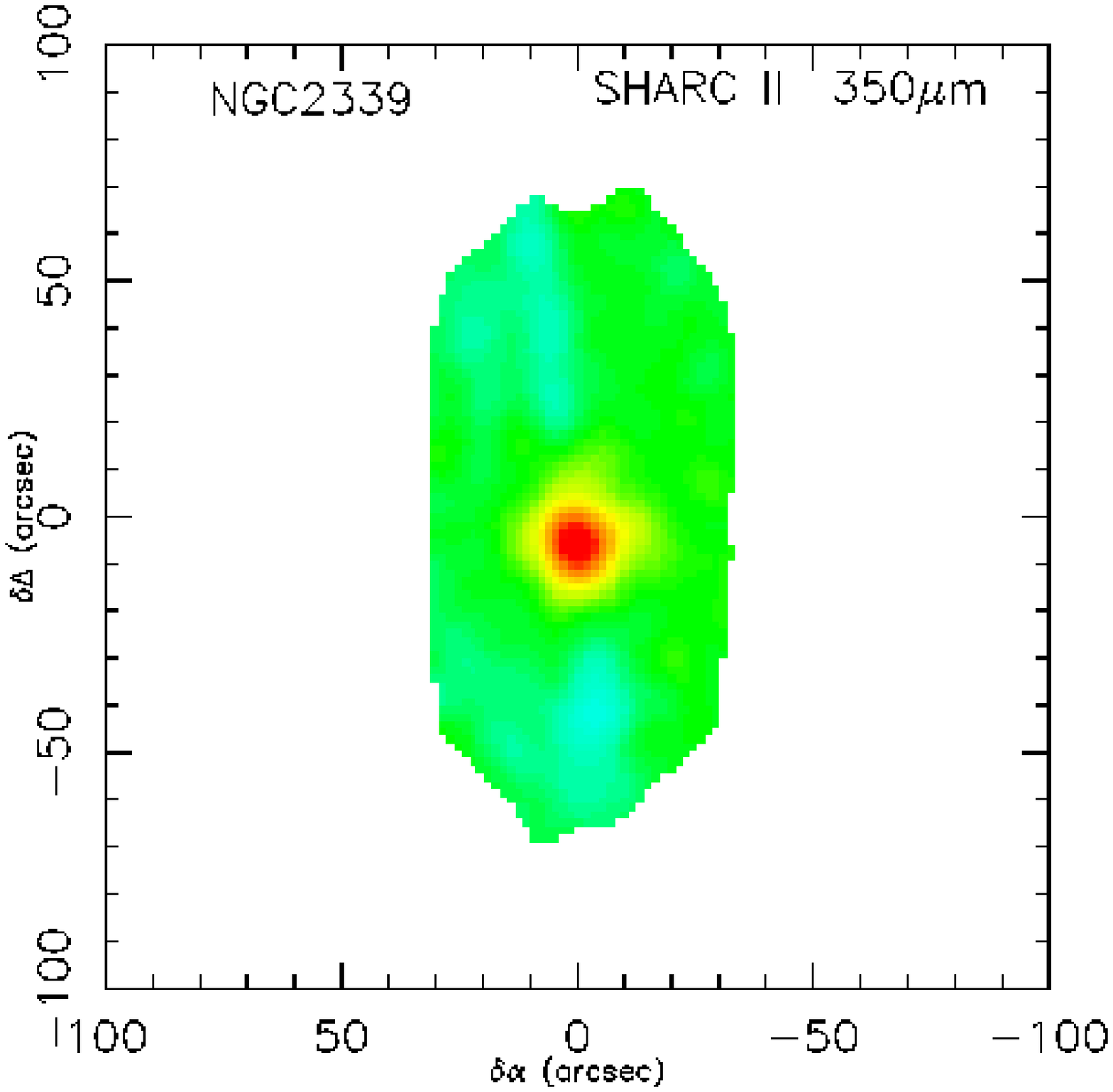}}&
{\includegraphics[width=1.5in, angle=270]{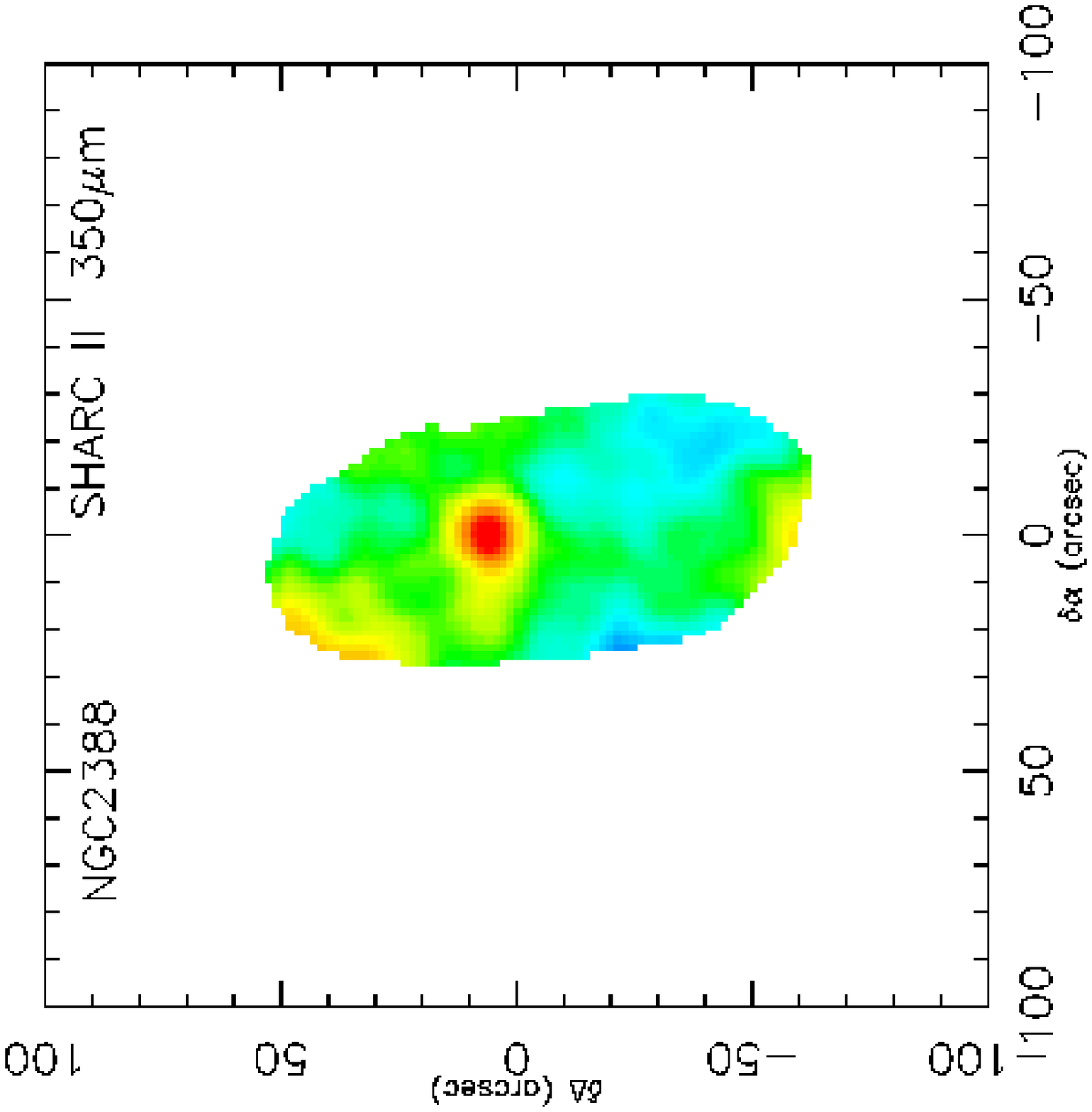}}\\
{\includegraphics[width=1.5in, angle=270]{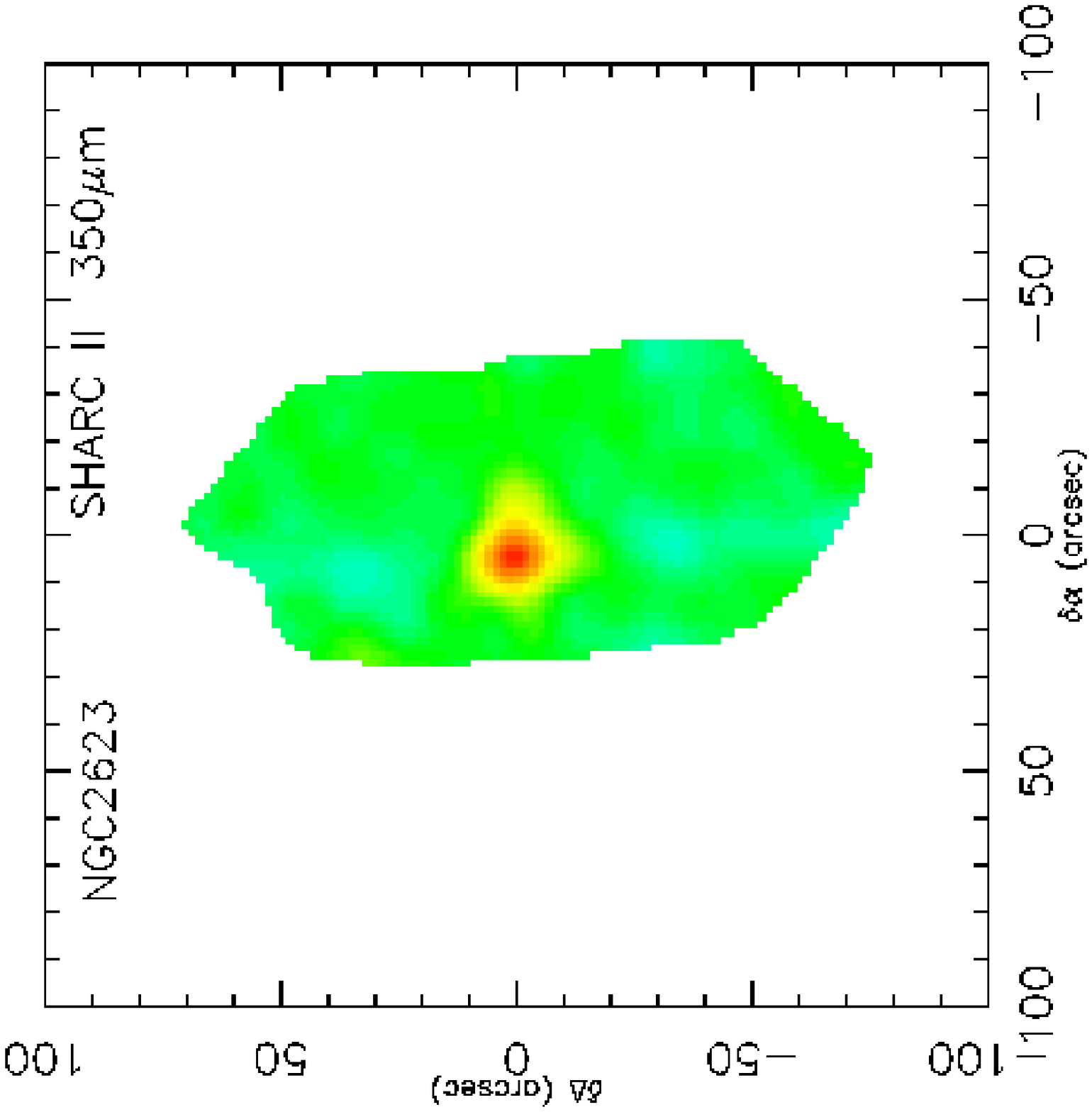}}&
{\includegraphics[width=1.5in, angle=270]{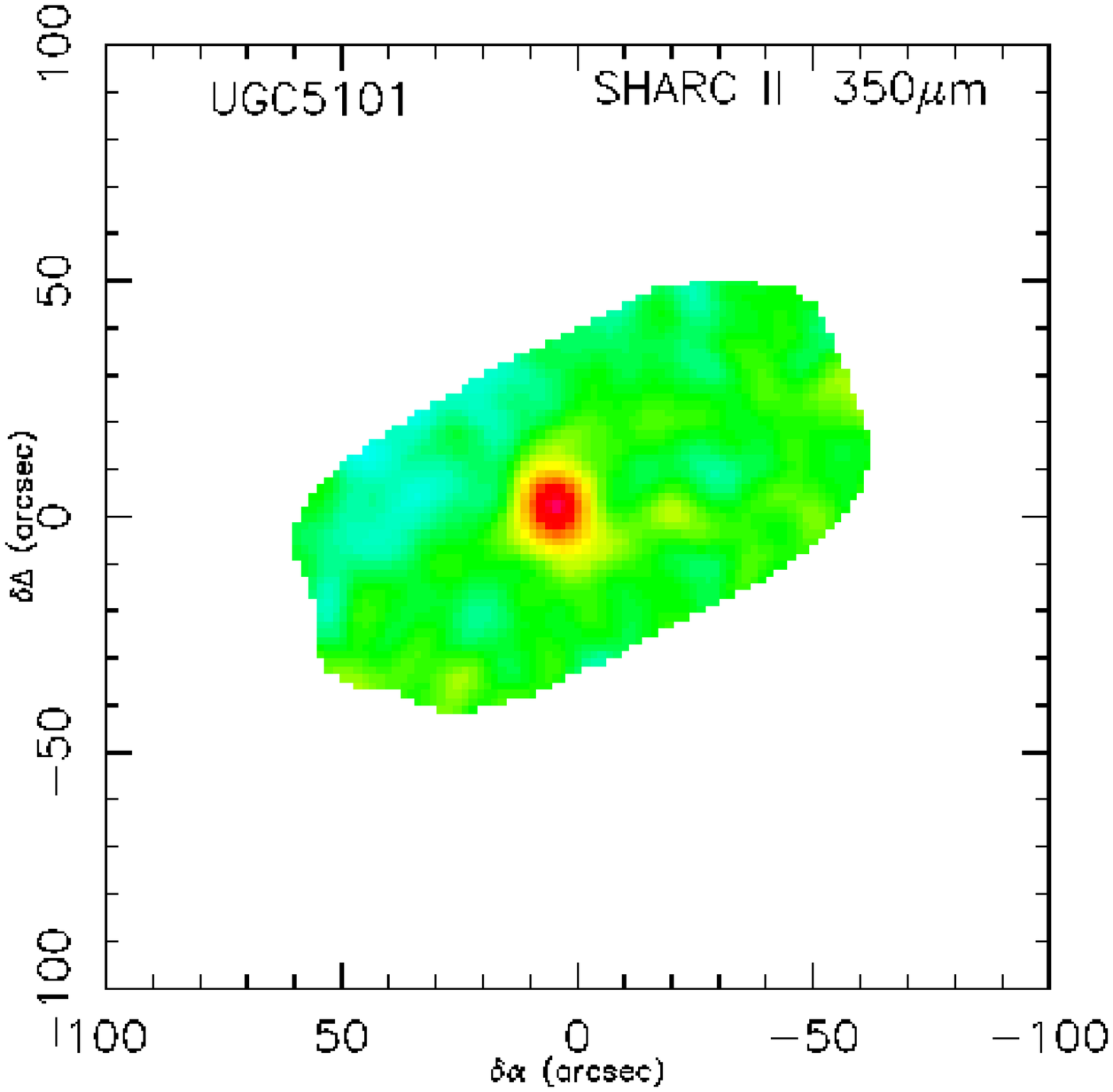}}&
{\includegraphics[width=1.5in, angle=270]{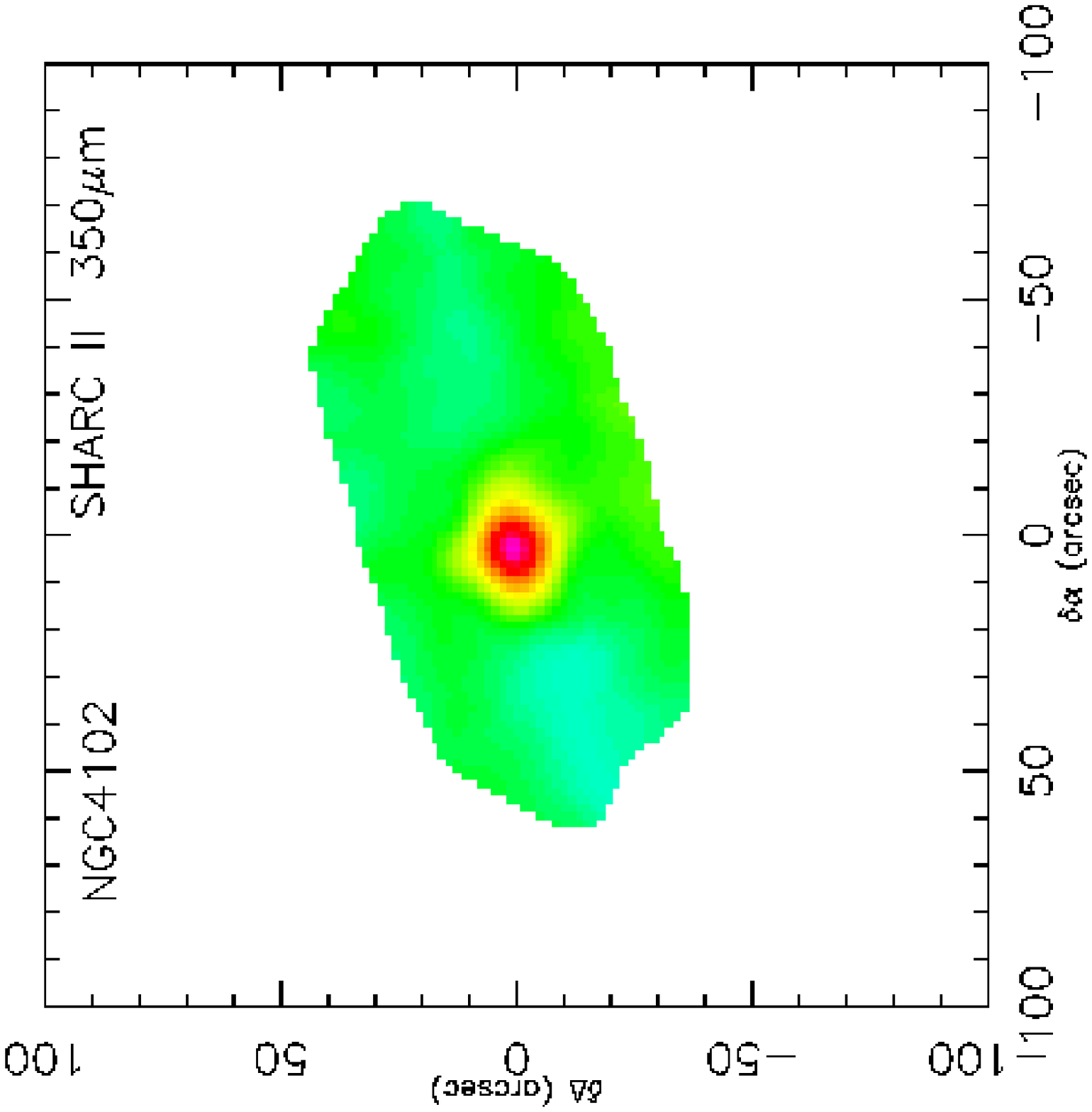}}&
{\includegraphics[width=1.5in, angle=270]{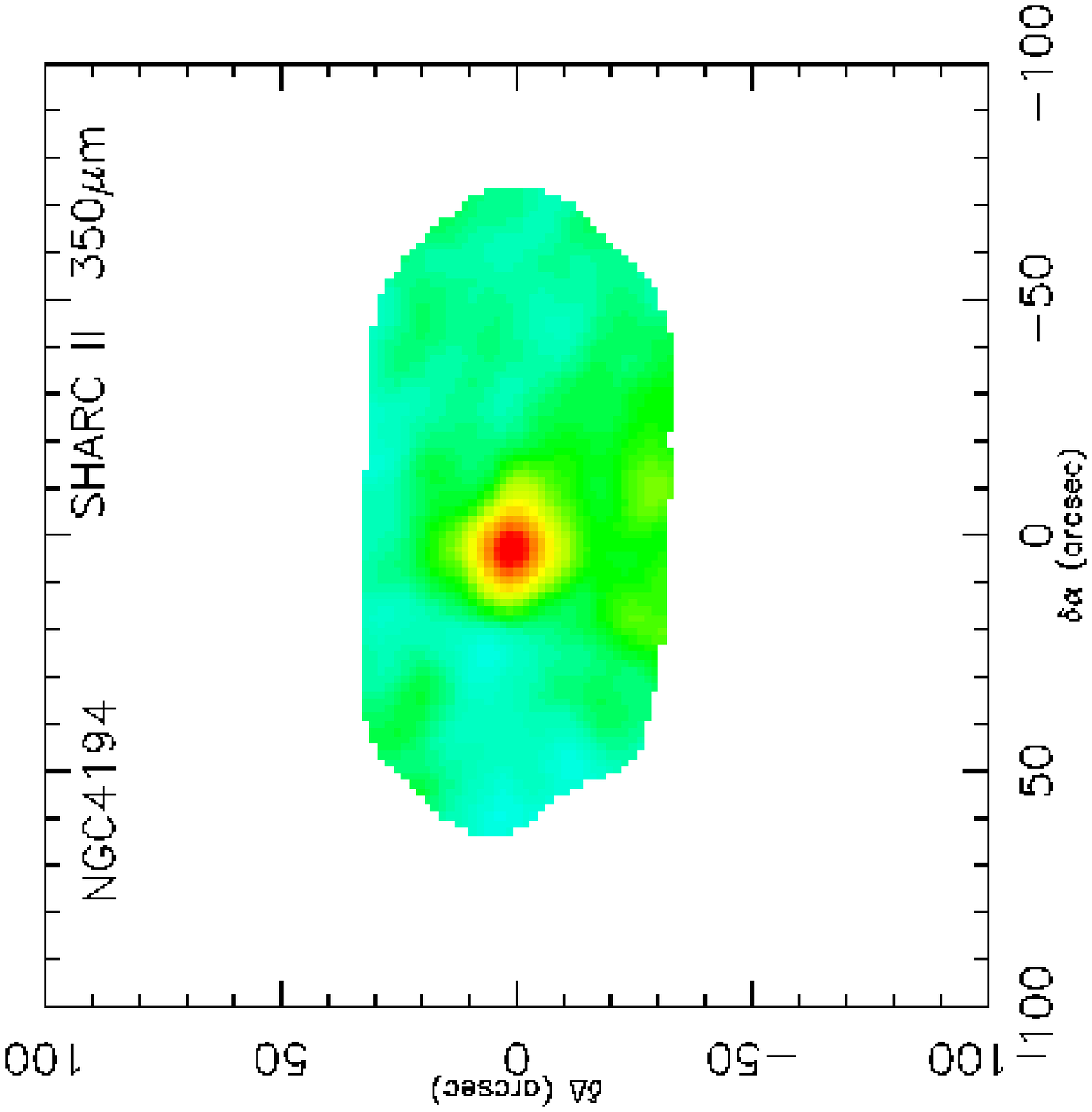}}\\
{\includegraphics[width=1.5in, angle=270]{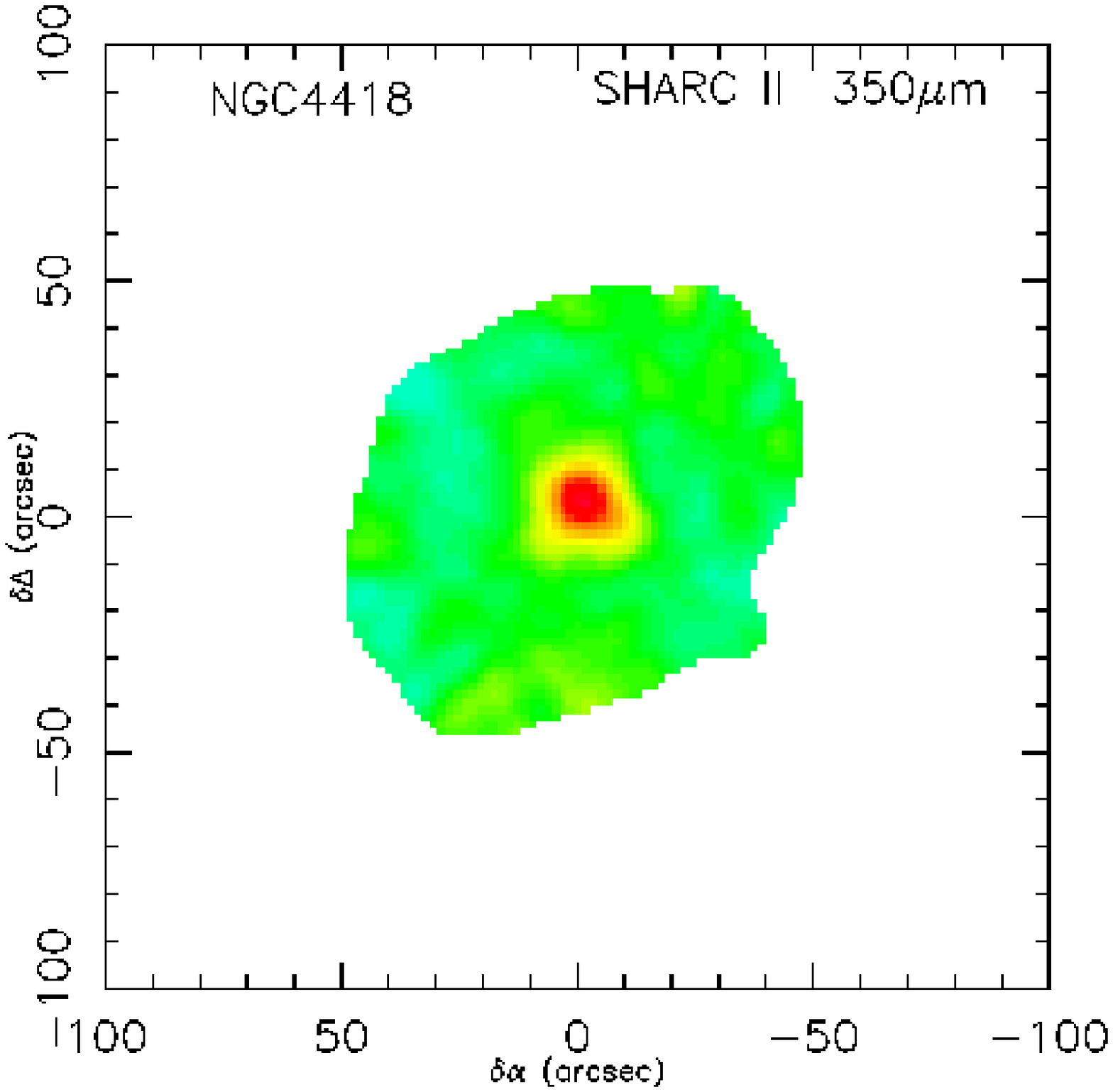}}&
{\includegraphics[width=1.5in, angle=270]{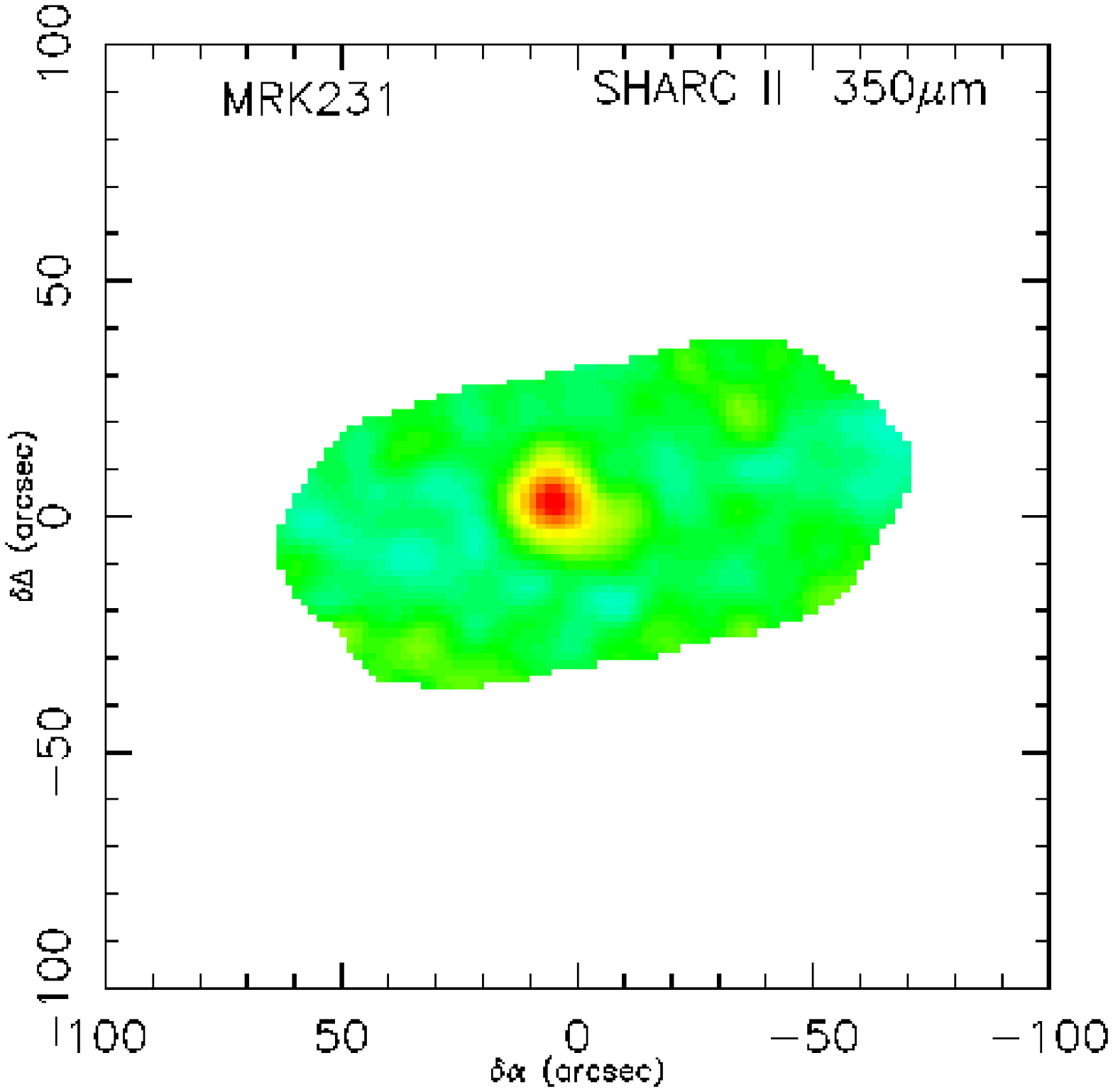}}&
{\includegraphics[width=1.5in, angle=270]{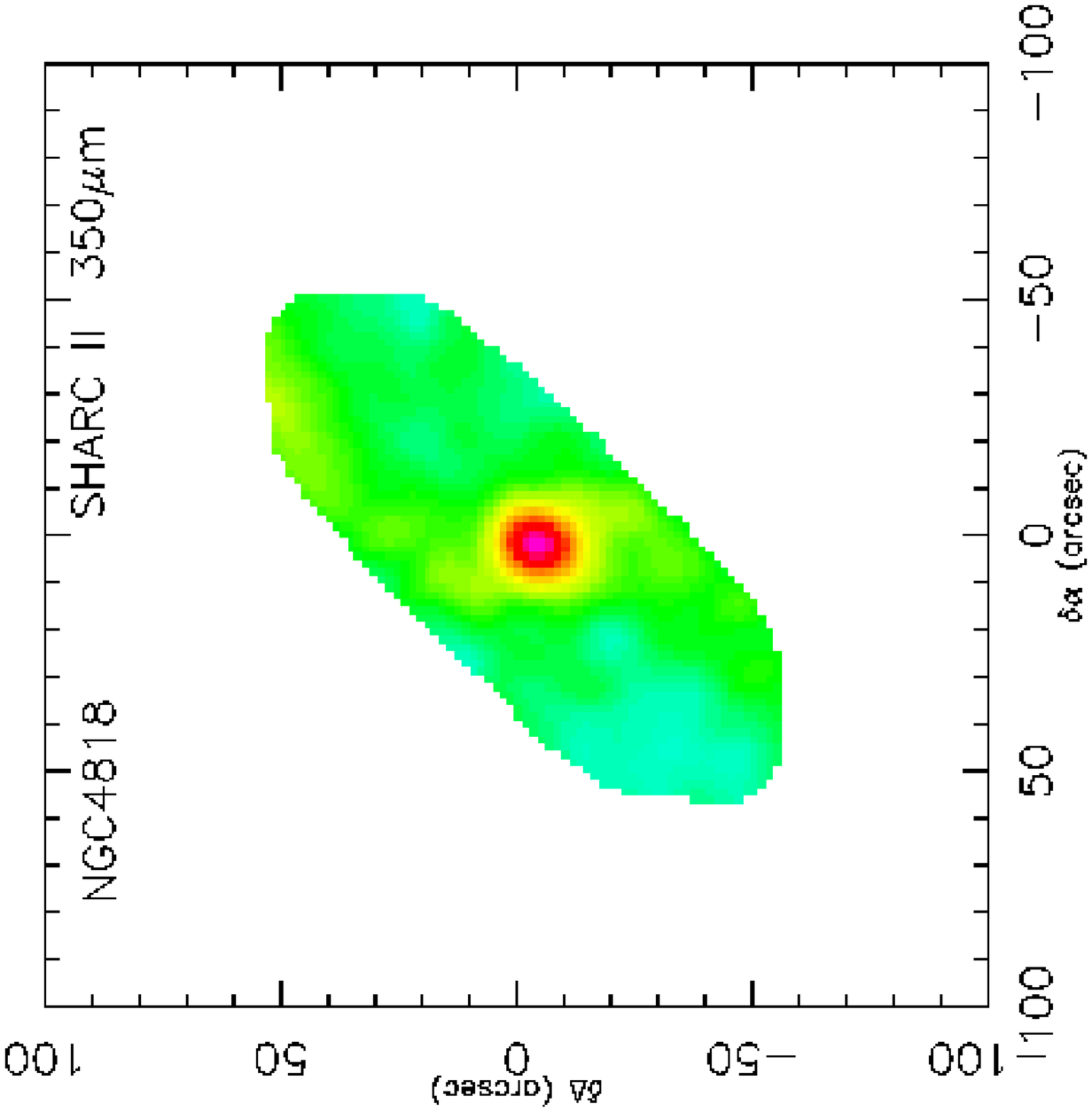}}&
{\includegraphics[width=1.5in, angle=270]{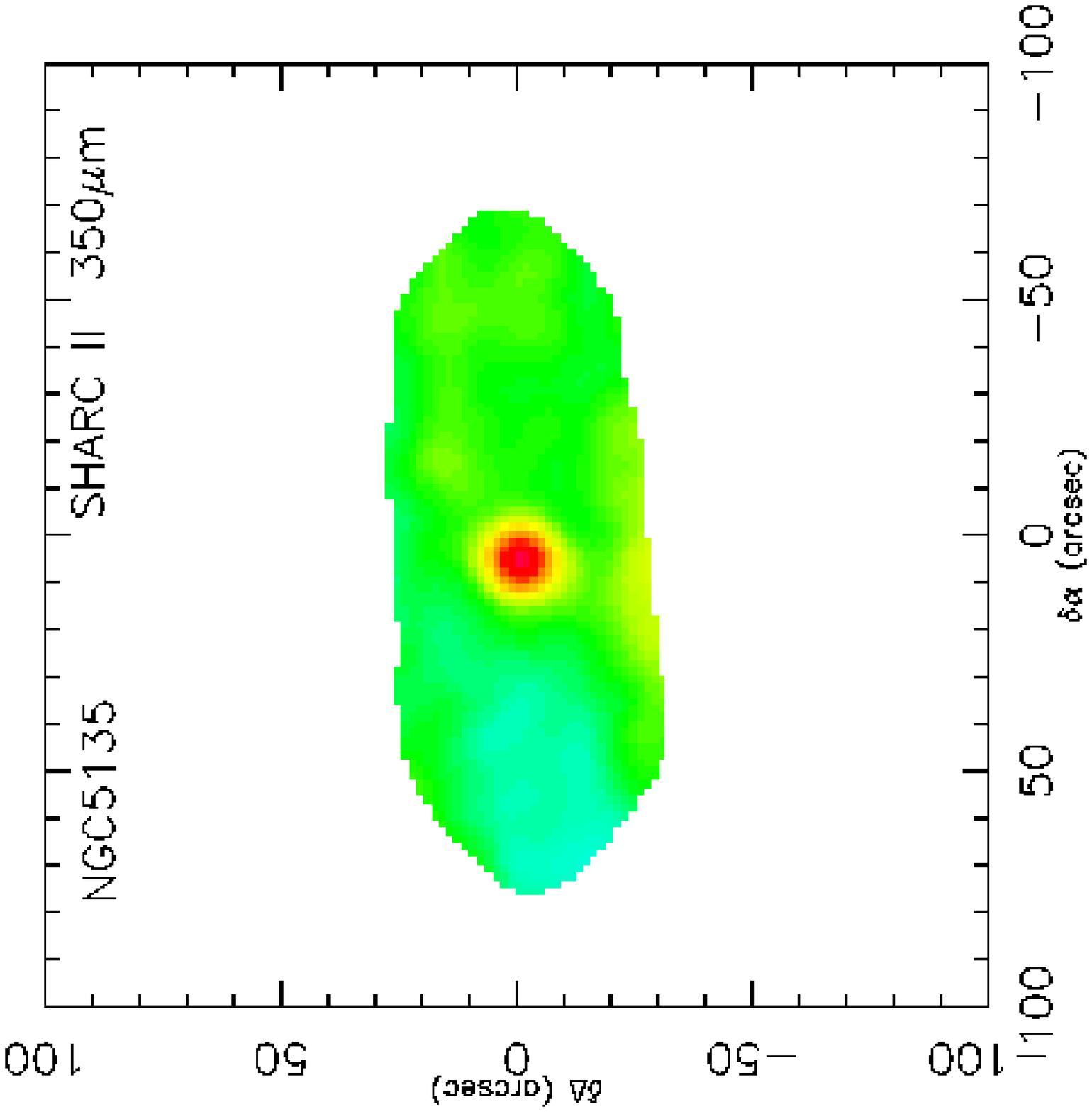}}\\
{\includegraphics[width=1.5in, angle=270]{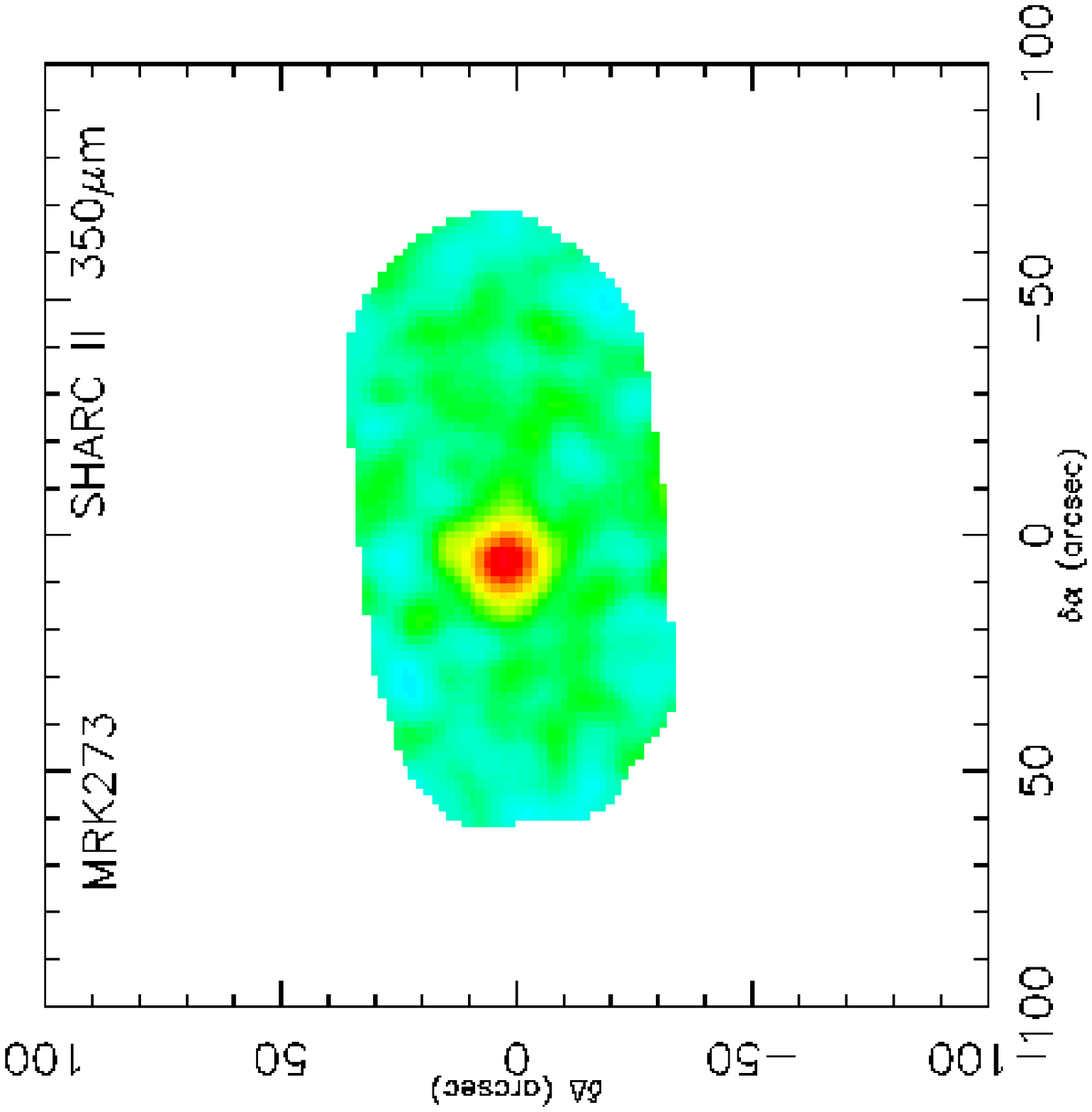}}&
{\includegraphics[width=1.5in, angle=270]{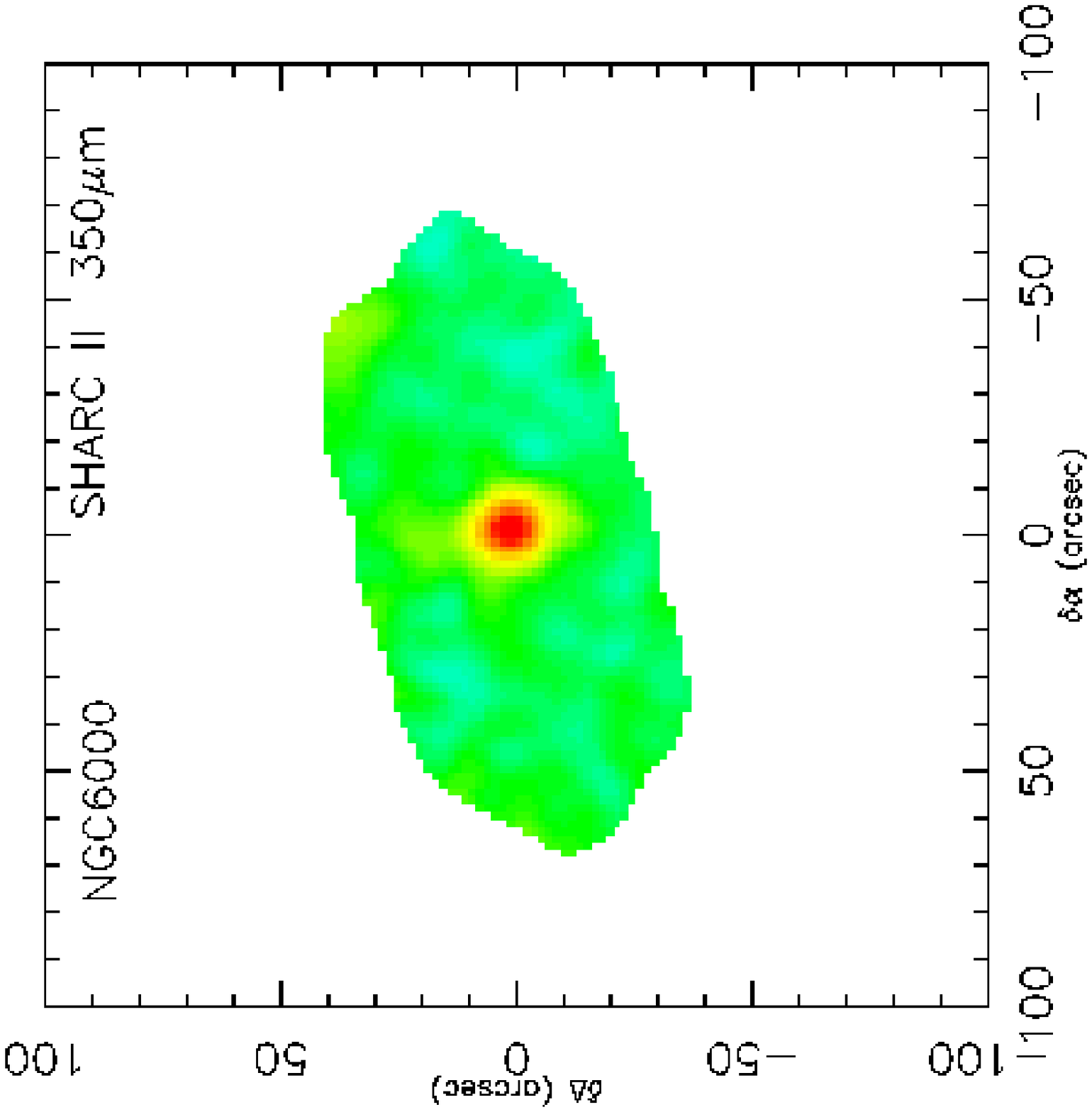}}&
{\includegraphics[width=1.5in, angle=270]{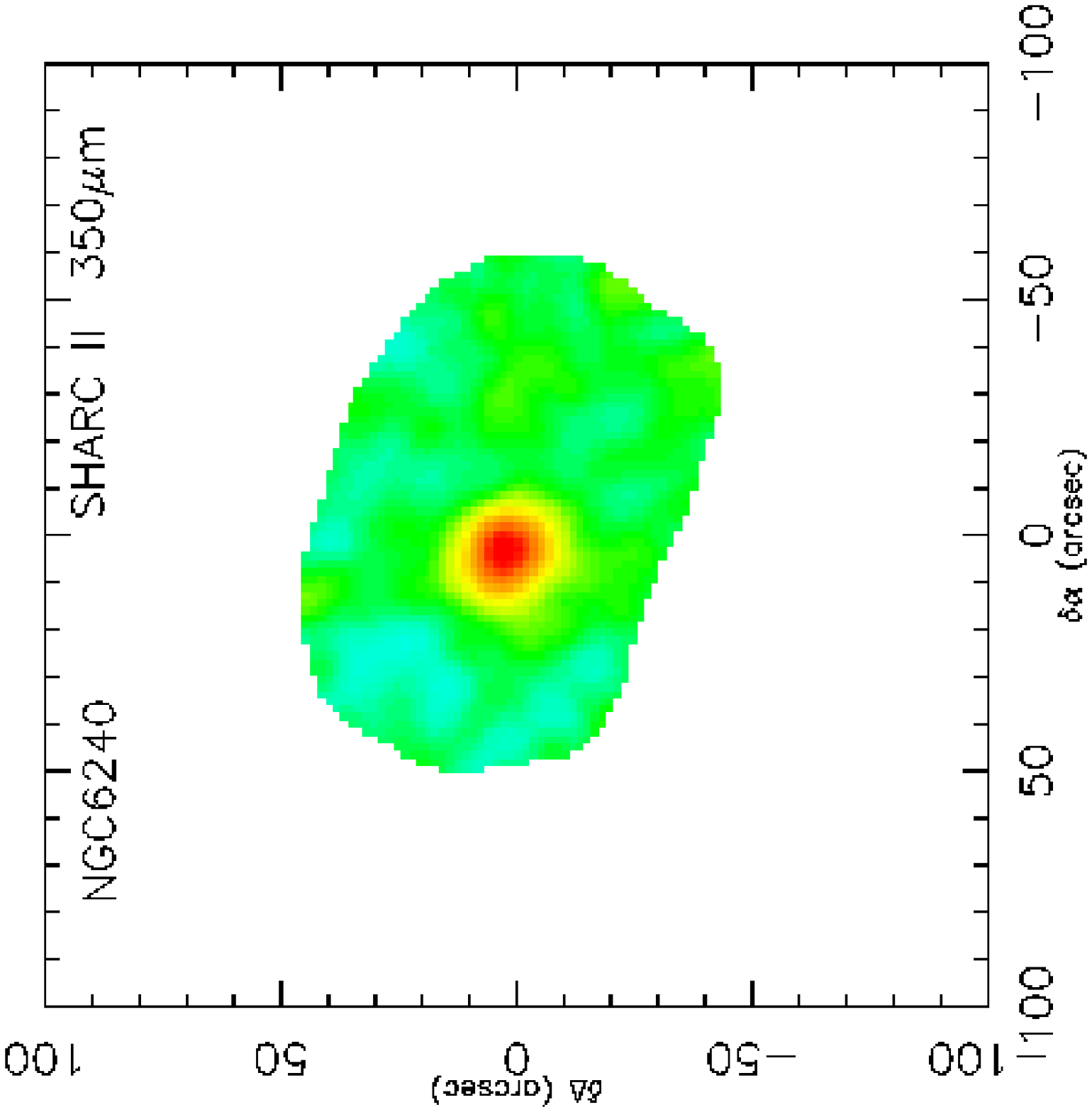}}&
{\includegraphics[width=1.5in, angle=270]{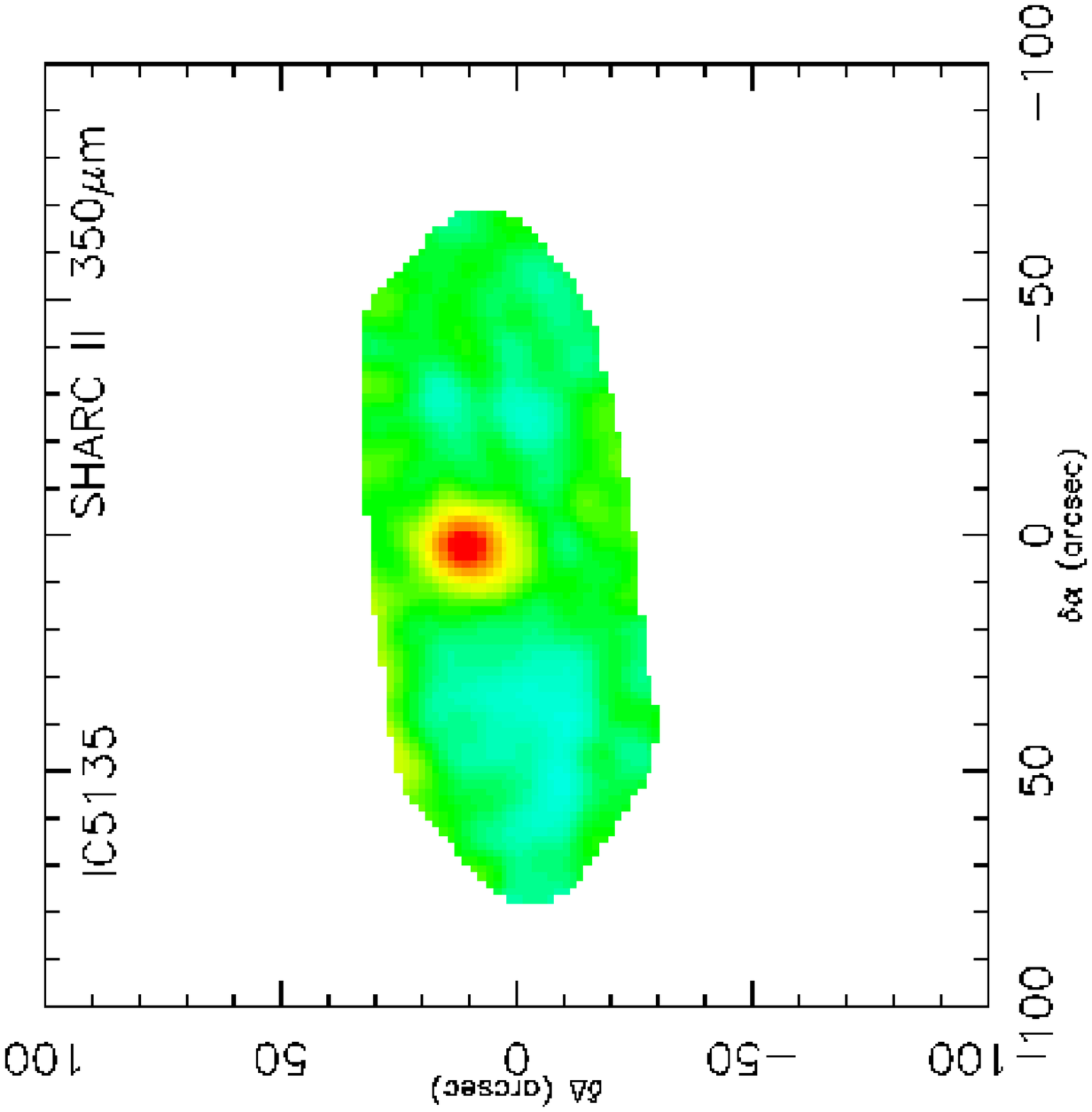}}\\
{\includegraphics[width=1.5in, angle=270]{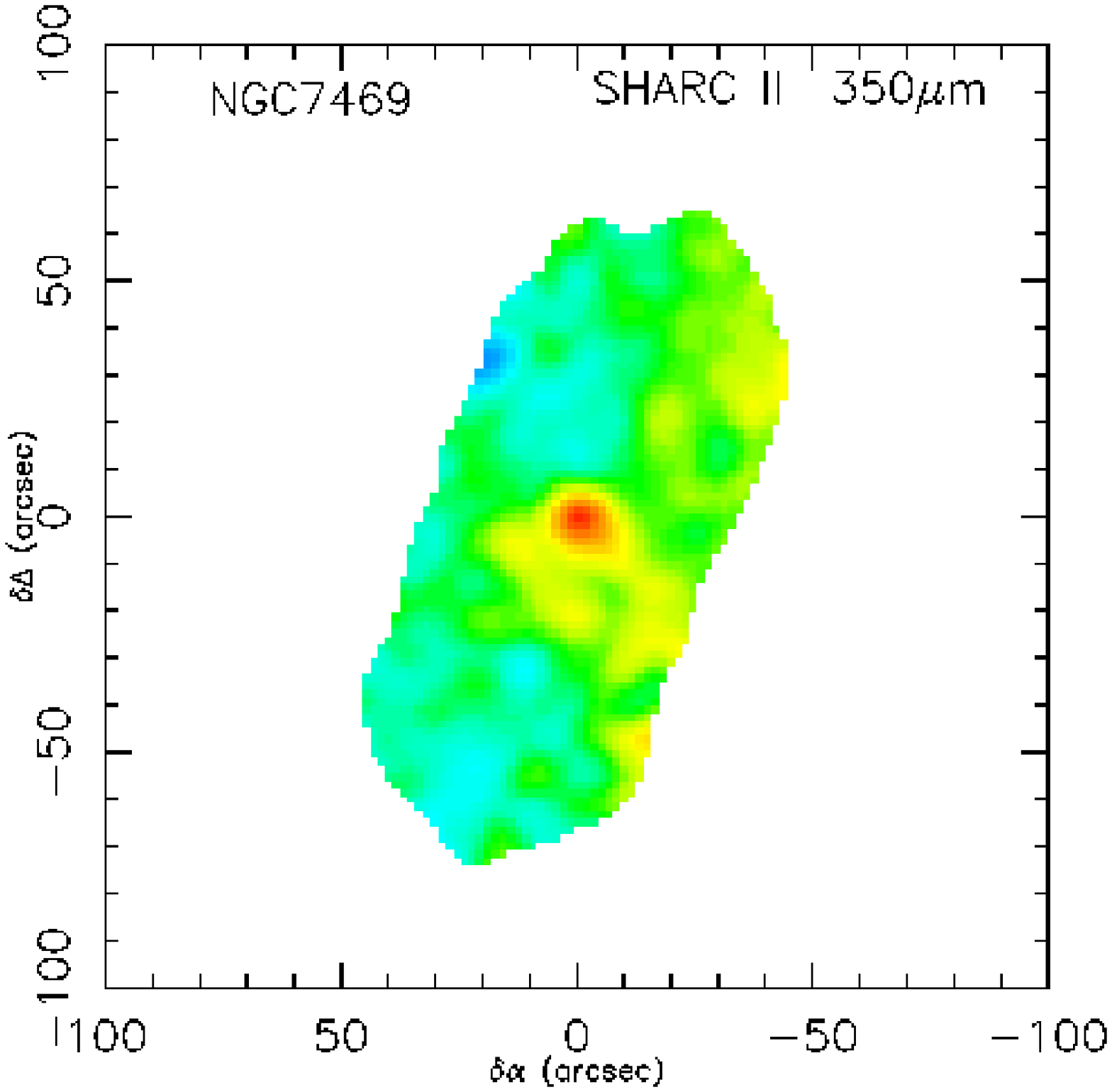}}&
{\includegraphics[width=1.5in, angle=270]{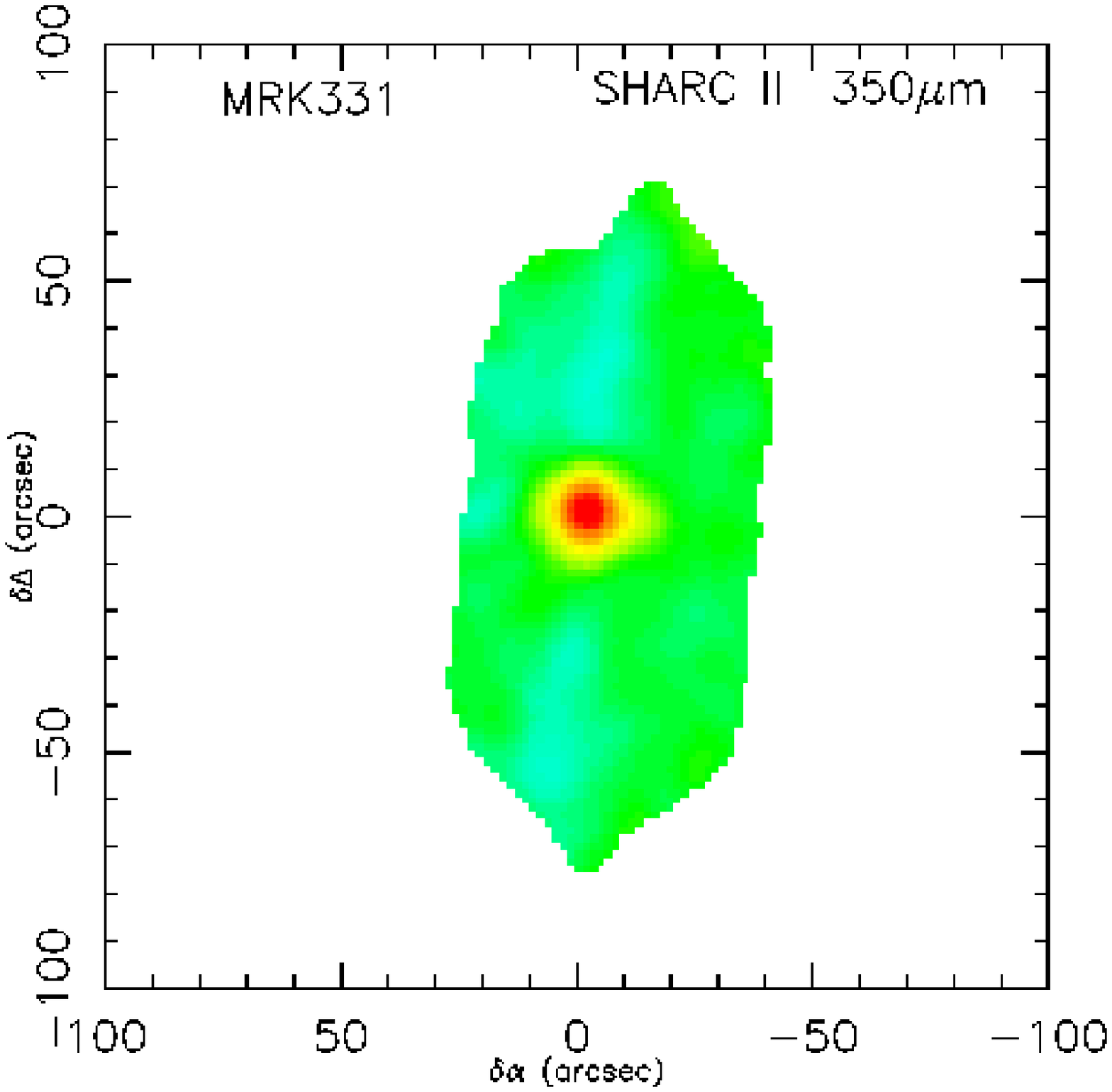}}\\
\end{tabular}
\caption{SHARC-II 350\um maps of galaxies in the local LIRG sample. 
In each map, the green plane represents the part of bolometer array 
which has at least half of the maximum exposure, and the red peak in 
the middle of the array indicates a detected source ($\rm S/N \geq 10$).}
\label{local-350map}
\end{center}
\end{figure}

The observations were made during a series of SHARC-II (Dowell et al.\ 2003)
observing runs from January 2003 to September 2004 at the CSO under 
moderate weather conditions ($\rm 0.07\leq \tau_{225\,GHz} \leq 0.10$). 
Integration time varied from source to source, 
depending on the brightness of each source at 350\ums, and the
atmospheric opacity and sky variability. On average, 0.5 hours 
of integration time was spent on each source. 
All measurements were made by scanning the bolometer array in a Lissajous pattern 
centered on each source; 
scans were taken without chopping the secondary mirror.
Pointing and calibration scans were taken on an hourly basis on strong submillimeter sources. 
The absolute calibration was found to be accurate within 20\% error margin. 
Raw data were reduced using the software package CRUSH (Kov\'{a}cs\ 2006),
to obtain a 350\um map and flux density for each source.
All of the 18 sources were clearly detected at levels of $\rm S/N \geq 10$, and 
Fig.\ \ref{local-350map} shows the SHARC-II 350\um maps
clipped to have at least half of the maximum exposure.
Table\ \ref{local-s350-sed} lists the 350\um flux densities and flux errors.

\section{Spectral Fits and Derived Quantities}\label{local-sed}
We adopt the single-temperature, optically-thin far-IR/submm SED model, as given by 
\mbox{Eq.\ (\ref{sed})}. A flux error at a uniform 20\% is assumed at each data point, 
hence each flux measurement is weighted equally in the nonlinear least squares 
fit. All but four galaxies 
(NGC\,2388, NGC\,4194, NGC\,4818 and NGC\,5135) in the 
sample have photometry data at far-IR/summ/mm wavelengths in addition to
60, 100 and 350\ums, thus the dust temperature and 
emissivity index can both be treated as free parameters. 
However, the SED fitting procedure introduces a significant degeneracy
between the estimated values of \Td and \bet (Blain et al.\ 2003) - a problem that persists
even when the frequency sampling of the far-IR/submm SED 
is highly redundant. To achieve redundancy in the cases of NGC\,2388, NGC\,4194, NGC\,4818 and NGC\,5135, 
where the sampling of the SED is limited to three wavelengths,
we assume $\beta=1.5$ - a value most commonly assumed and observed.

\begin{figure}
\begin{center}
{\includegraphics[width=4in, angle=90]{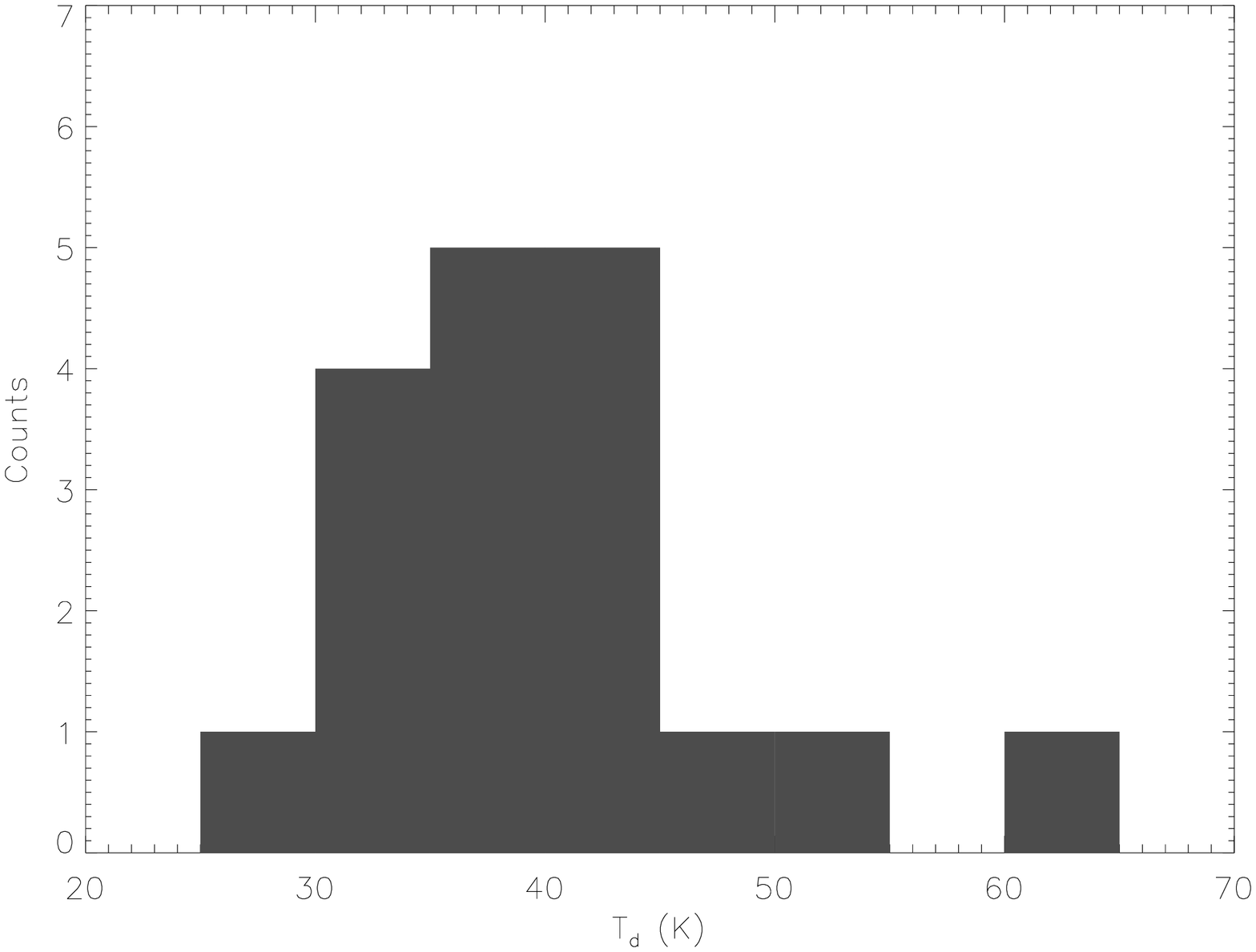}}
{\includegraphics[width=4in, angle=90]{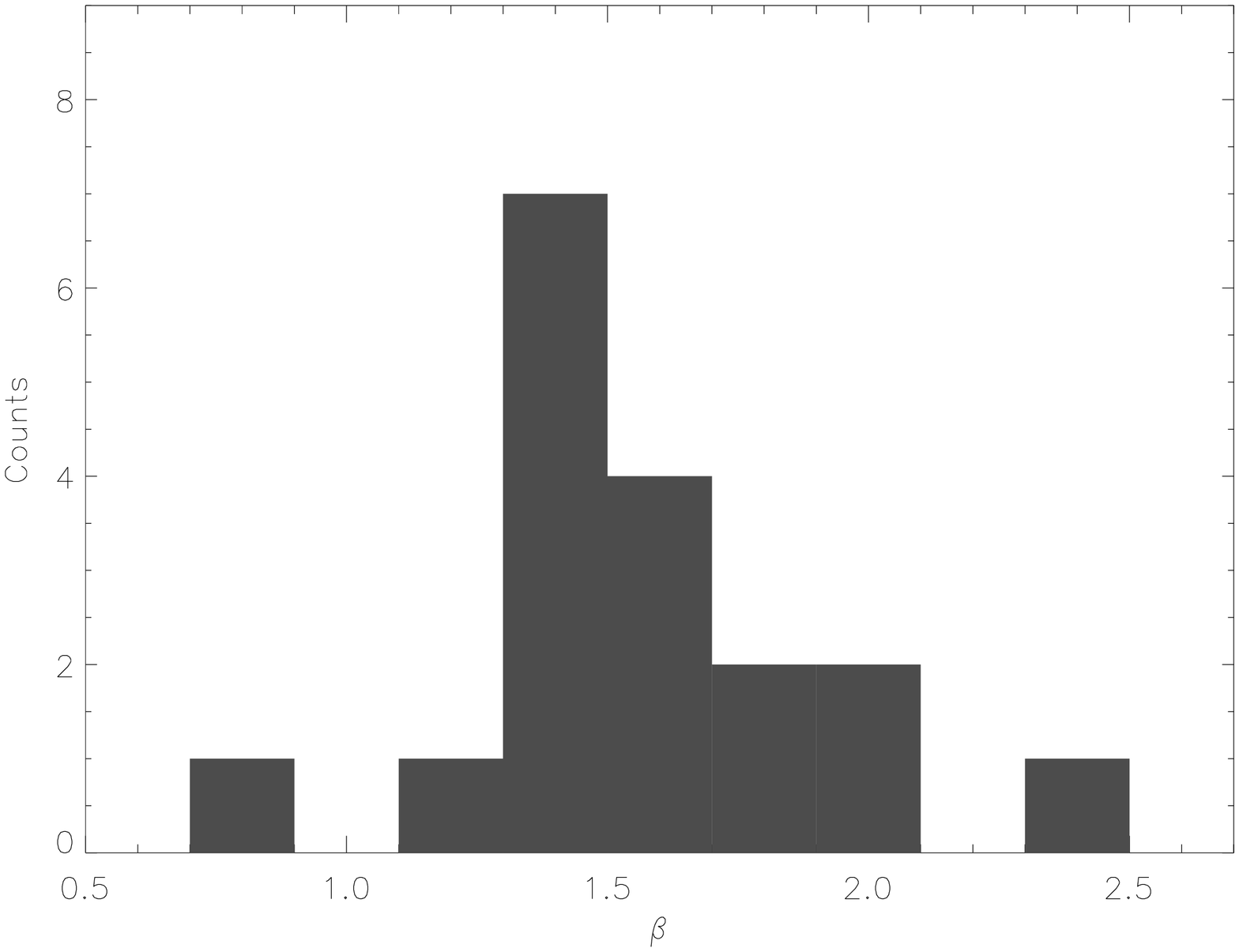}}
\caption{Histograms of \Td and \bet derived for the local LIRG sample.}\label{local-histo-para}
\end{center}
\end{figure}

The best-fitting dust temperature has a 
sample median and standard deviation of $\rm T_d = 39.4 \pm 7.9 \,\rm K$
and ranges within $\rm 29.2 \leq T_d \leq 62.9$\Ks, 
with two galaxies (NGC\,4418 and Mrk\,231) showing $\rm T_d \geq 50\,K$. 
The emissivity index ranges within $\rm 0.9 \leq \beta \leq 2.4$, and 
has a sample median of $\rm \beta = 1.6 \pm 0.3$. 
The far-IR/submm SED fits of our local LIRG sample are shown in Fig.\ \ref{local-sedplot}; 
the distributions of \Td and \bet are shown in Fig.\ \ref{local-histo-para}. 

The global far-IR luminosity (\Lfirs) and dust mass (\Mds) are calculated from
\begin{eqnarray}
\rm L_{\mbox{\tiny{FIR}}} & = & \rm 4 \pi \, D_{\tiny{L}}^{\,2} \, \int_{40  \, \mu m}^{1000  \, \mu m} S_\nu \, d\nu \label{lfir} \\
\rm M_{d} & = & \rm \frac{S_\nu \, D_{\tiny{L}}^{\,2}}{\kappa_{\nu} \, B_\nu(T_d)} \label{md}
\,,\end{eqnarray}
where \DL is the luminosity distance, and $\rm \kappa_{\nu} \equiv \frac{3 \, Q_{\nu}}{4 \, a \, \rho} $ 
is the dust mass absorption coefficient, varying as $\rm \propto \nu^{\,\beta}$.
We adopt dust parameters given by Hildebrand (1983): 
$ \rm Q_{125{\mu}m}=7.5\times10^{-4}$, $\rm \rho = 3.0\,g/{cm^3}$ and $a \rm = 0.1\,{\mu}m$, which implies
$\kappa_{125 \mu m} = 1.875 \, \rm (kg/{m^2})^{-1}$.
While there is significant uncertainty associated with $\kappa$,
due to the lack of accurate knowledge of interstellar dust properties, 
\Lfir is tightly constrained when a satisfactory SED fitting is achieved. 
The \Lfir and \Md values inferred for our local LIRG sample range within 
$\rm 10^{\,9.9} \leq L_{\mbox{\tiny{FIR}}} \leq 10^{\,12.2}$ \Lsun
and $\rm 10^{\,6.1} \leq M_d \leq 10^{\,\,8.3}$ \Msuns, and
have sample medians of 
$\rm L_{\mbox{\tiny{FIR}}} = 10^{\,11.2 \pm 0.6} \,L_\odot$ and $\rm M_d = 10^{\,7.3 \pm 0.6} \,M_\odot$. 
Thus, our sample spans more than two orders of magnitude in both 
far-IR luminosity and dust mass. 
Table\ \ref{local-s350-sed} lists the fitted values of \Td and \bets, 
along with the \Lfir and \Md estimates for our local LIRG sample.

\section{\mbox{\Tds-\bet} Relation}\label{local-tb}

\subsection{Correlation Coefficient \& Simulations}\label{tb_degeneracy}
A negative correlation between the dust temperature and emissivity index
clearly emerges from the \mbox{\Tds-\bet} scatter plot for our local LIRG 
sample (Fig.\ \ref{tb-scatter}). 
Note that, in the investigation of the \mbox{\Tds-\bet} relation, 
we only include the 14 sources for which the frequency 
sampling is sufficient ($\geq4$) to allow simultaneous estimates of \Td and \bets. 
The non-parametric correlation coefficient
\footnote{Without prior knowledge of the true distributions of dust temperature and emissivity index,
the non-parametric (rank) correlation coefficient 
between these two variables provides a more robust correlation test than the parametric 
correlation coefficient (Press\ 1999).}
between the dust temperature and emissivity index 
estimated for our 14 local LIRGs is calculated to have a value of $\rm \rho_{np} = -0.79$, 
and it is non-zero at the significance level of $(1-p)$, where $p \sim 7.43 \times 10^{-4}$ (Yang\ 2006). 
We thus conclude the inverse \mbox{\Tds-\bet} correlation observed in 
our local LIRG sample is statistically significant.

However, we note that fitted values of \Td and \bet are simultaneously obtained from the 
SED fitting, in which a strong and negative \mbox{\Tds-\bet} degeneracy is present 
and could potentially introduce an artificial inverse \mbox{\Tds-\bet} correlation.
We address this problem by repeatedly performing SED fittings on simulated 
photometric data. Our simulation method, described as follows, is a modification 
and extension of the simulation procedure performed by Dupac et al.\ (2002).
First, we are primarily concerned with the non-parametric \mbox{\Tds-\bet} 
correlation coefficient; second, the effects of the \mbox{\Tds-\bet} degeneracy 
on the observed \mbox{\Tds-\bet} correlation can be accurately quantified.

Two sets of independent, uniformly distributed random numbers are generated within the 
ranges of $20-60$ and $1.0-2.5$, and 
they are taken to be the simulated values of \Td and \bet respectively 
for each of the 14 local LIRGs. 
Furthermore, we assume flux errors to be normally distributed
and at a constant level of 20\% at all wavelengths.
Using the simulated values of \Tds, \bet and measurement errors, 
simulated photometric data are calculated at all far-IR/submm wavelengths 
where observed data are available for 
each of the 14 sources in our local LIRG sample. 
SED fitting of the simulated photometric data for each source follows, 
yielding fitted values of \Td and \bets. 
We calculate non-parametric correlation coefficients 
between the ``real'' (simulated) as well as 
the fitted values of \Td and \bet in the sample. 

\begin{figure}
\begin{center}
{\includegraphics[width=5in, angle=90]{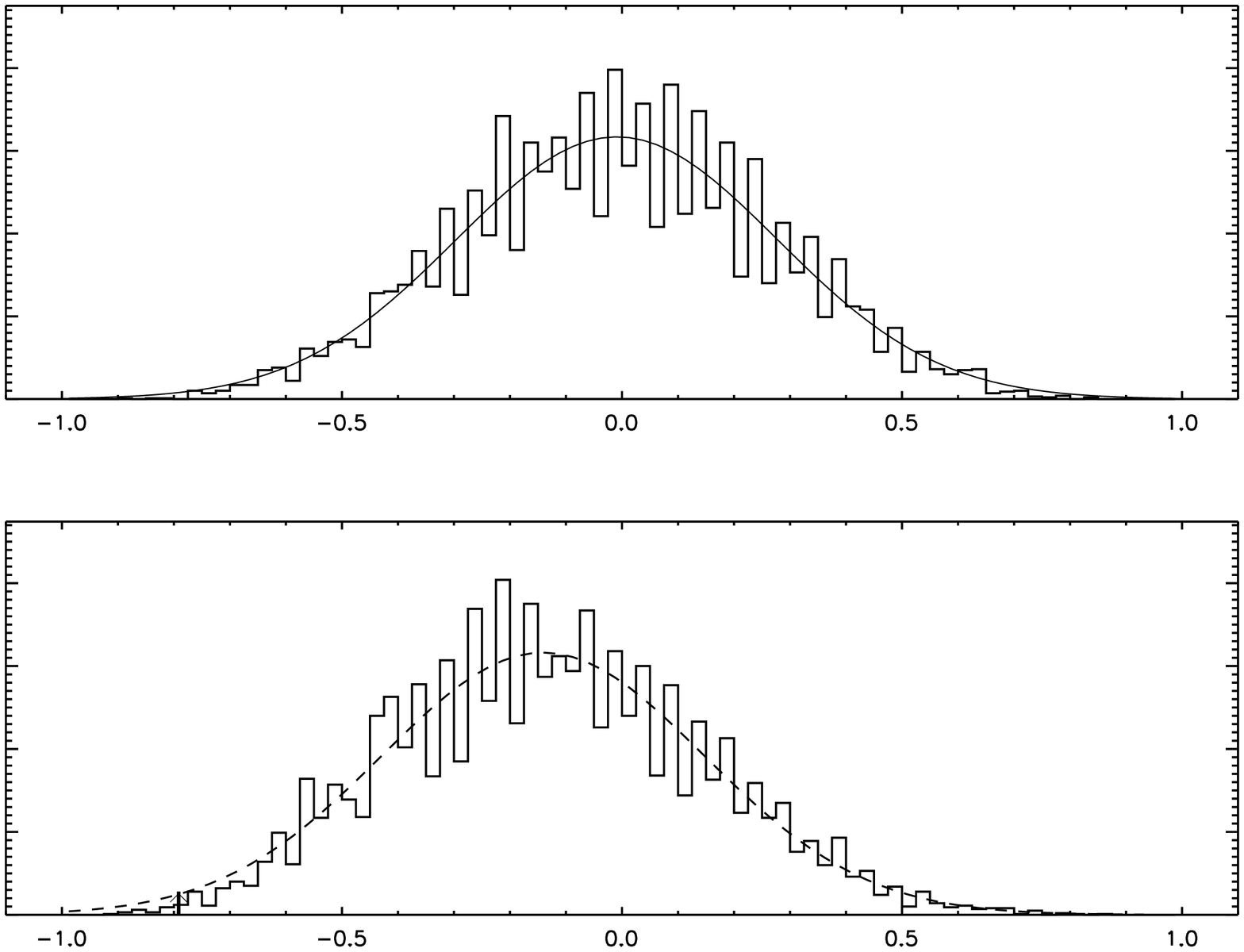}}
\caption{{\bf a):} The distribution of $\rm \rho_{np}$ calculated for
the uncorrelated, simulated \Tds-\bet is well approximated by a Gaussian function 
$\sim \rm N(-0.01, 0.29)$, as shown by the solid line.
{\bf b):} The distribution of $\rm \rho_{np}$ calculated for
the fitted \Tds-\bets, from SED fittings of the simulated 
photometric data, is well approximated by a Gaussian function $\sim \rm N(-0.14, 0.29)$,
as shown by the dashed line. Under this distribution, the $\rm \rho_{np}$ of the fitted \Tds-\bets,
from SED fittings of the observed photometric data, is clearly significant 
at a value of $\rm \rho_{np} = -0.79$ (marked by the sticker).}\label{tb_corr_sim}
\end{center}
\end{figure}

The above steps are repeated many times ($ \sim 10000$), and 
the probability distribution function of the 
non-parametric correlation coefficient between the dust temperature and emissivity index
is thereby established. As shown in Fig.\ \ref{tb_corr_sim},  
$\rm \rho_{np}$ of the simulated \Tds-\bet show a distribution 
well approximated by a Gaussian function 
$\sim {\rm N}(-0.01, 0.29)$, as can be expected from the central limit theorem given
that the simulated \Td and \bet are independent random variables. By contrast,
$\rm \rho_{np}$ of the fitted \Tds-\bet roughly follows a Gaussian
distribution $\sim {\rm N}(-0.14, 0.29)$, suggesting 
the SED fitting procedure typically introduces a non-parametric \mbox{\Tds-\bet} 
correlation coefficient at the level of 
$\rm \rho_{np} \sim -0.14$ when \Td and \bet are in fact uncorrelated,
given the SED frequency samplings of our 14 local LIRGs in the far-IR/submm/mm.
Under the normal distribution $\sim {\rm N}(-0.14, 0.29)$, 
the observed \mbox{\Tds-\bet} correlation coefficient, 
$\rm \rho_{np} = -0.79$, 
is significant at the level of $(1-p^\ast)$, where 
$p^\ast \sim  1.20 \times 10^{-2}$. 
Hence, we deem the negative
\mbox{\Tds-\bet} degeneracy in the SED fitting insufficient 
to explain the inverse \mbox{\Tds-\bet} correlation observed 
in the local LIRG sample and conclude that the 
observed inverse \mbox{\Tds-\bet} correlation is real.

\subsection{Functional Fitting}
\begin{figure}
\begin{center}
{\includegraphics[width=5in, angle=90]{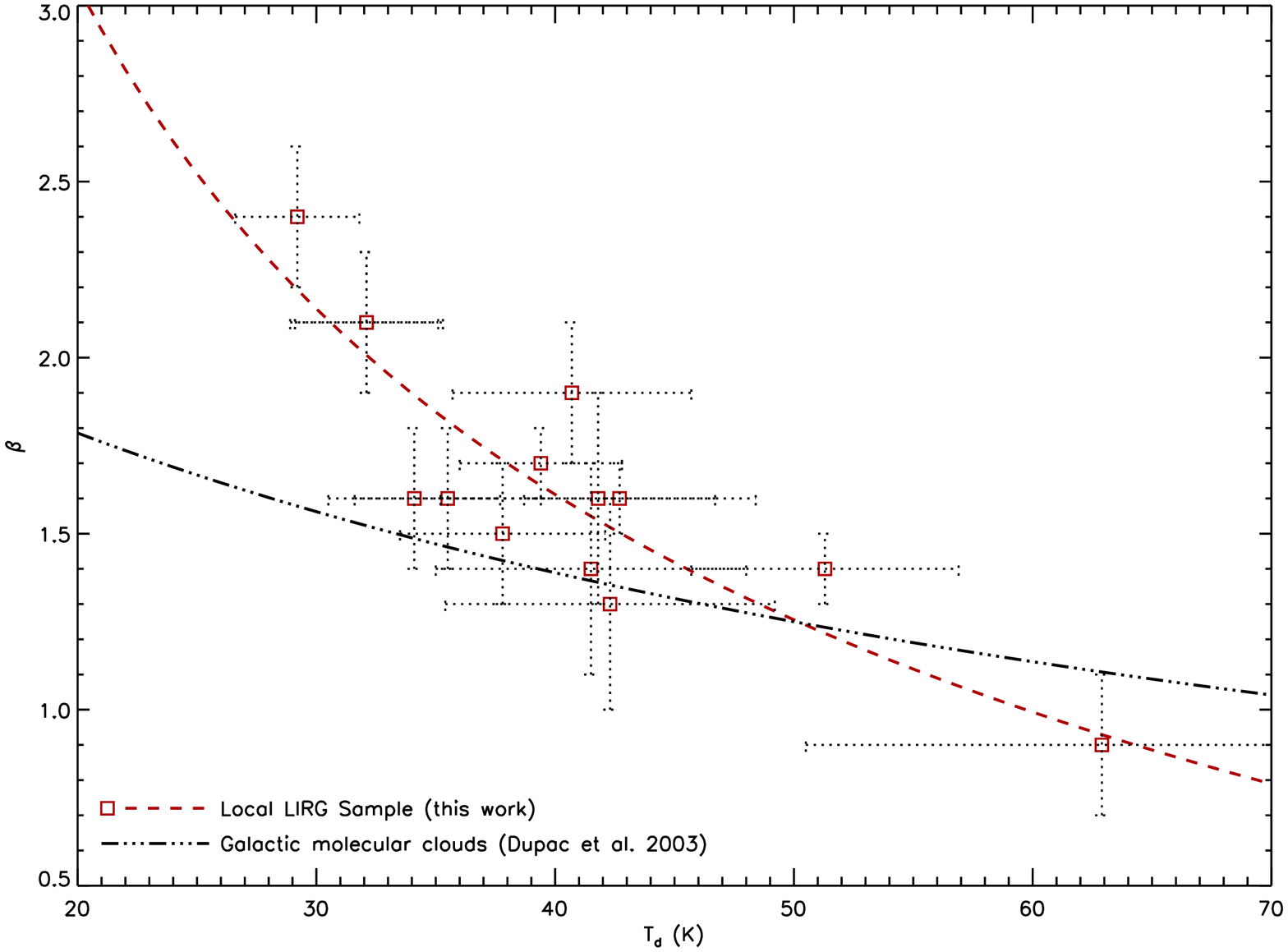}}
\caption{The inverse \mbox{\Tds-\bet} correlation observed in our local LIRG sample
when the far-IR/submm/mm SEDs are sampled at no less than four wavelengths.
The dashed line represents the best fit to the \Td and \bet values derived for the 14 local LIRGs. 
For comparison, the \mbox{\Tds-\bet} correlation derived for the Galactic molecular cloud sample 
(Dupac et al.\ 2003) is shown by the dashed-dotted line.}\label{tb-scatter}
\end{center}
\end{figure}

Based on physical considerations, particularly Eq.\ (\ref{equi_td1}), 
we fit the inverse \mbox{\Tds-\bet} correlation observed in our local LIRG sample
using the function
\begin{eqnarray}
\rm T_d & = & c_1 \, ^ {\frac{1}{c_2 + \beta}} \label{tb-function-a}
\,.\end{eqnarray}
The best fit to Eq.\ (\ref{tb-function-a}) is
\begin{eqnarray}
\rm T_d & = & [9.86 \times 10^{9}] \, ^ {\frac{1}{4.63 + \beta}} \label{tb-fit-a}
\,,\end{eqnarray}
which provides a good fit to the dust temperatures and emissivity indices 
estimated for our 14 local LIRGs,
as shown in Fig.\ \ref{tb-scatter}.
We also try two alternative functional forms in which the dust emissivity index 
is a power or hyperbolic function of the dust temperature, i.e. 
\begin{eqnarray}
\beta & = & c_1 \, {\rm T_d} \, ^{c_2}           \label{tb-function-b}\\
\beta & = & \frac{1}{c_1 + c_2\,\rm T_d}       \label{tb-function-c}
\,.\end{eqnarray}
The best fits to Eqs.\ (\ref{tb-function-b}) and (\ref{tb-function-c}) are
$\rm \beta  =  [82.37] \, T_d \, ^{-1.07}$ 
and 
$\rm \beta  =  [62.89]/{T_d}$
respectively, and yield \mbox{\Tds-\bet} curves nearly identical to 
Eq.\ (\ref{tb-fit-a}). Thus, we consider Eq.\ (\ref{tb-fit-a}) 
as the best description of the \mbox{\Tds-\bet} 
relation observed in the local LIRG sample. 
This \mbox{\Tds-\bet} relation is also tested in the SED fittings 
of a ULIRG sample at intermediate redshifts ($0.1 \ls z \ls 1.0$, Yang\ 2006), 
for which the incorporation of the \mbox{\Tds-\bet} relation (Eq.\ \ref{tb-fit-a})
leads to reasonable SED fits for the vast majority of the sources. 
This provides strong support for an extension of 
the \mbox{\Td-\bet} relation, as derived for our local LIRG sample, 
to dusty galaxies in the more distant Universe.

For comparative purpose, we also plot the \mbox{\Tds-\bet} relation 
given by Dupac et al.\ (2003), as derived from multiband submm
observations of a large sample of molecular clouds in the Galaxy (Fig.\ \ref{tb-scatter}). 
These authors reported a wide range of \Td and \bet values estimated 
from SED fittings and found an inverse \mbox{\Tds-\bet} 
correlation best fitted by a hyperbolic function
$\beta= \frac{1}{0.4 + 0.08\,\rm T_d}$.
Clearly, the \mbox{\Tds-\bet} correlation seen in the Galactic molecular clouds
is significantly different from, in fact much flatter than, that observed 
in our local LIRG sample, despite the agreement in qualitative trend. 
We note that the \Td and \bet estimates for the molecular clouds 
are sometimes given by SED fittings without flux measurements at wavelengths shortward 
of the SED peaks, which would otherwise have greatly improved the accuracy of 
the \Td estimates. The difference in the shape of the observed \mbox{\Tds-\bet} correlations, 
as derived for astronomical samples of differing nature, scales and environments,
is intriguing. We argue that variations in the \mbox{\Tds-\bet} relation 
are not unreasonable, given that various physical characteristics, 
such as the radiation field, grain properties and grain distributions, 
could all impact the large-scale dust temperature and emissivity index, 
and thus the way in which \Td and \bet relate to each other.
There is considerable observational challenge currently
facing us in the task of achieving sufficient far-IR/submm/mm SED samplings, 
for a large number of dusty objects 
over a wide range in redshift, physical scale and environment.
Dedicated observational efforts in the future 
would allow us to more precisely quantify the 
\mbox{\Tds-\bet} relation, discover its full range and investigate its evolution.

\subsection{Laboratory Results}\label{lab}
Laboratory measurements of the absorption coefficients of interstellar 
grain analogs at submm/mm wavelengths have revealed an intrinsic 
temperature dependence of grain absorption coefficient and emissivity index
(Agladze et al.\ 1996; Mennella et al.\ 1998).
Agladze et al.\ (1996) measured absorption spectra of crystalline and 
amorphous grains at wavelengths ranging between 700\um and 2.9\mm at 
temperatures between 1.2 and 30\Ks. The emissivity index was found to range 
between 1.2 and 2.5, and show a negative temperature dependence 
for amorphous grains at temperatures above 10\Ks.
Mennella et al.\ (1998) measured absorption coefficients of crystalline 
and amorphous dust analog grains between 20\um and 2\mm over 
the temperature range 24$-$295\Ks. The temperature and wavelength domains 
covered in this set of laboratory experiments are probably more relevant for 
understanding astronomical observations at far-IR/submm wavelengths.
The emissivity index, as well as its temperature dependence, was found to be 
sensitive to grain material and grain structure, 
and variations in the absorption coefficient and emissivity index
were found to be more prominent in amorphous grains than in crystalline grains. 
These authors found a monotonic increase of the absorption coefficient 
with dust temperature, while the emissivity index was found to range from 
0.5 to 2.3 and decrease as the dust temperature increases.

\subsection{Explanations}
We conclude the inverse \mbox{\Tds-\bet} correlation observed in our local LIRG sample is 
statistically significant and physically real, and offer three possible explanations.

First, Eq.\ (\ref{equi_td1}) tells us that 
the equilibrium dust temperature displays a negative dependence on the emissivity index, 
for a given radiation field and grain material.

Second, the grain emissivity index has intrinsic negative temperature dependence, 
as evidenced by laboratory measurements (\S~\ref{lab}). This effect
can be understood by considering the Schl\"{o}mann (1964) model 
which treats an amorphous solid as an ionic crystal with a random charge distribution.  
Within this framework, the absorption spectrum is dominated by photon-to-phonon 
coupling at all frequencies. We argue that the observed negative temperature 
dependence of grain emissivity index is attributed to the decreasing 
grain dimension ($D$) at higher temperatures (i.e., grain structure becomes more ``open'') 
\footnote{Here we adopt the concept of 
fractal dimension (Hausdorff dimension), defined as 
\mbox{$D \rm{\equiv log_{P/p}{N}}$}, where P/p is the size ratio of 
self-similar fractal objects and N is the number of smaller units 
that can fit into the larger unit. Structures with lower fractal dimension are more ``open''.},
which results in a flatter frequency dependence of the phonon mode density
and consequently lower emissivity indices at higher dust temperatures (Yang\ 2006).

Third, an inverse \mbox{\Tds-\bet} relation can be caused by mixtures of non-uniform 
grain populations, particularly those involving very small grains
or grains covered with ice mantles. \bet values as high as 3.5 have been predicted
for dust grains covered with ice mantles (Aannestad 1975), whereas very small grains 
are expected to have $\beta \sim 1$ (Seki \& Yamamoto\ 1980). 
At the same time, low dust temperatures 
favor formation of ice mantles and lead to strong emission at longer wavelengths, 
while high dust temperatures (or temperature fluctuations) in very small grains prohibit 
accretion of ice mantles and give rise to strong emission at shorter wavelengths. 
As a result, the dust temperature and emissivity index observed in the large-scale tend to 
move in the direction of low-\Tds-high-\bet when the overall dust emission 
is dominated by that of cold dust possibly covered with ice mantles,
while \Td and \bet shift in the direction of high-\Tds-low-\bet when very small grains dominate the overall emission.

\subsection{Applications}
A \mbox{\Tds-\bet} relation, if established over a wide range of dusty galaxies,
would be extremely useful for modeling their observed far-IR/submm SEDs, as 
it effectively reduces the number of variables (therefore the 
number of required flux measurements) by one in the SED fitting. 
This effect is especially desirable for studying faint, distant submm 
sources and interpreting results from single submm wavelength 
imaging and deep field surveys, given the considerable 
observational challenge typically faced by deep submm observations.
For a given photometric dataset, a \mbox{\Tds-\bet} relation  
either allows the estimate of one more parameter in the SED fitting, 
or improves the uncertainties associated with the fitted variables, as
exemplified by the SED modeling of a sample of intermediate-redshift ULIRGs 
(Yang\ 2006), for which the sampling of the far-IR/submm SED is limited 
to three wavelengths. By using the \mbox{\Tds-\bet} relation, 
both the dust temperature and emissivity index can be fitted, 
which represents a major advance over 
the more conventional and commonly utilized SED modeling method, 
which, without the application of a \mbox{\Tds-\bet} relation, 
would only fit for one of the two very important parameters.

\section{Conclusion}
We have observed and detected 18 local LIRGs ($0.003 \leq z \leq 0.042$) 
at 350\ums, using the SHARC-II camera at the CSO. The acquired 350\um data, 
in combination with pre-existing far-IR/submm/mm photometric
data, lead to accurate estimates of SED properties of these galaxies. 
We find sample medians of dust temperature (\mbox{$\rm T_d = 39.4 \pm 7.9 \,\rm K$}), 
emissivity index (\mbox{$\rm \beta = 1.6 \pm 0.3$}),
far-IR luminosity (\mbox{$\rm L_{\mbox{\tiny{FIR}}} = 10^{\,11.2 \pm 0.6} \,L_\odot$})
and dust mass (\mbox{$\rm M_d = 10^{\,7.3 \pm 0.6} \,M_\odot$}).
An inverse \mbox{\Tds-\bet} correlation, 
best described by \mbox{$\rm T_d  =  [9.86 \times 10^{9}] \, ^ {\frac{1}{4.63 + \beta}}$}, 
is established for our local LIRG sample. 
This effect is most likely caused by intrinsic interdependence 
between the dust temperature and emissivity index 
as physical parameters, as revealed by laboratory measurements (\S~\ref{lab}), 
as as well as variations in grain properties in the ISM. 
The establishment of a \mbox{\Tds-\bet} relation is extremely useful 
in the far-IR/submm SED fittings of dusty galaxies, as it serves to
effectively reduce the number of free parameters by one in the SED 
function.

\acknowledgments
We would like to thank J. Zmuidzinas, A. Blain, T. Soifer, T. Greve and C. Dowell for 
helpful discussions. We are grateful to the referee for improving this paper through 
thoughtful comments. The CSO is supported by the NSF fund under contract AST 02-29008.

\begin{singlespace}
\bibliographystyle{apj}

\begin{thebibliography}{}
\bibitem[Aannestad(1975)]{aan75}Aannestad P., 1975, \apj, 200, 30
\bibitem[Agladze et al.(1996)]{agl96}Agladze N., Sievers A., Jones S., Burlitch J., Beckwith S., 1996, \apj, 462, 1026
\bibitem[Benford (1999)]{ben99}Benford D., 1999, Caltech Ph.D. Thesis
\bibitem[Blain et al.(2003)]{bla03}Blain A., Barnard V., Chapman S., 2003, \mnras, 338, 733
\bibitem[Blake et al.(1996)]{bla96}Blake G., Mundy L., Carlstrom J., Padin S., Scott S., Scoville N., Woody D., 1996, \apj, 472, L49
\bibitem[Bohren \& Huffman(1983)]{boh83}Bohren C., Huffman D., 1983, \emph{Absorption and Scattering of Light by Small Particles}
\bibitem[Dowell et al.(2003)]{dow03}Dowell C., Allen C., Babu R., Freund M., Gardner M., Groseth J., Jhabvala M., Kov\'{a}cs A., Lis D., Moseley S., Phillips T., Silverberg R., Voellmer G., Yoshida H., 2003, SPIE, 4855, 73
\bibitem[Dunne et al.(2000)]{dun00}Dunne L., Eales S., Edmunds M., Ivison R., Alexander P., Clements D., 2000, \mnras, 315, 115
\bibitem[Dupac et al.(2003)]{dup03}Dupac X., Bernard J., Boudet N., Giard M., Lamarre J., M\'{e}ny C., Pajot F., Ristorcelli I., Serra G., Stepnik B., Torre J., 2003, \aap, 404, L11
\bibitem[Greenberg(1978)]{gre78}Greenberg J., 1978, in McDonnell J. (ed.), \emph{Cosmic Dust}
\bibitem[Hildebrand(1983)]{hil83}Hildebrand R., 1983, \qjras, 24, 267
\bibitem[Kittel(1976)]{kit76}Kittel C., 1976, \emph{Introduction to Solid State Physics}
\bibitem[Knapp et al.(1993)]{kna93}Knapp G., Sandell G., Robson E., 1993, \apjs, 88, 173
\bibitem[Kov\'{a}cs(2006)]{kov06thesis}Kov\'{a}cs A., 2006, Caltech Ph.D. Thesis
\bibitem[Lis et al.(1998)]{lis98}Lis D., Serabyn E., Keene J., Dowell C., Benford D., Phillips T., Hunter T., Wang N., 1998, \apj, 509, 299
\bibitem[Mennella et al.(1998)]{men98}Mennella V., Brucato J., Colangeli L., Palumbo P., Rotundi A., Bussoletti E., 1998, \apj, 496, 1058
\bibitem[Press(1999)]{pre99}Press W., 1999, \emph{Numerical Recipes}
\bibitem[Roche \& Chandler(1993)]{roc93}Roche P., Chandler C., 1993, \mnras, 265, 486
\bibitem[Schwartz(1982)]{sch82}Schwartz P., 1982, \apj, 252, 589
\bibitem[Schl\"{o}mann(1964)]{sch64}Schl\"{o}mann E., 1964, \phrva, 135, 413
\bibitem[Seki \& Yamamoto(1980)]{sek80}Seki J., Yamamoto T., 1980, Ap\&SS, 72, 79
\bibitem[Wynn-William \& Becklin(1993)]{wwb93}Wynn-William C., Becklin E., 1993, \apj, 412, 535
\bibitem[Yang(2006)]{yang06thesis}Yang M., 2006, Caltech Ph.D. Thesis
\end{thebibliography}

\end{singlespace}

\begin{deluxetable}{lccccc}
\footnotesize
\tablecaption{The local LIRG sample selected for SHARC-II observations\label{local-target}}
\tablewidth{0pt}
\tablehead{\colhead{Source Name} & \colhead{RA} & \colhead{Dec} & \colhead{$z$} & \colhead{$\rm S_{60}$} & \colhead{$\rm S_{100}$} \\
\colhead{} & \colhead{J2000}  & \colhead{J2000} & \colhead{}  & \colhead{Jy} & \colhead{Jy}}
\startdata
NGC\,520  & 01h24m35.1s & +03d47m33s & 0.007609 & 30.37 & 46.15 \\
NGC\,1614 & 04h33m59.8s & -08d34m44s & 0.015938 & 33.02 & 34.35 \\
NGC\,2339 & 07h08m20.5s & +18d46m49s & 0.007358 & 18.45 & 31.46 \\
NGC\,2388 & 07h28m53.4s & +33d49m09s & 0.013790 & 17.03 & 25.33 \\
NGC\,2623 & 08h38m24.1s & +25d45m17s & 0.018463 & 23.80 & 26.66 \\
UGC\,5101 & 09h35m51.6s & +61d21m11s & 0.039390 & 12.24 & 20.25 \\
NGC\,4102 & 12h06m23.1s & +52d42m39s & 0.002823 & 47.04 & 73.84 \\
NGC\,4194 & 12h14m09.6s & +54d31m36s & 0.008359 & 24.09 & 26.06 \\
NGC\,4418 & 12h26m54.6s & -00d52m39s & 0.007268 & 45.58 & 31.99 \\
Mrk\,231  & 12h56m14.2s & +56d52m25s & 0.042170 & 33.28 & 30.29 \\
NGC\,4818 & 12h56m48.9s & -08d31m31s & 0.003552 & 20.32 & 26.72 \\
NGC\,5135 & 13h25m44.0s & -29d50m01s & 0.013716 & 16.93 & 30.72 \\
Mrk\,273  & 13h44m42.1s & +55d53m13s & 0.037780 & 23.09 & 21.97 \\
NGC\,6000 & 15h49m49.5s & -29d23m13s & 0.007315 & 36.48 & 51.76 \\
NGC\,6240 & 16h52m58.9s & +02d24m03s & 0.024480 & 22.87 & 27.21 \\
IC\,5135  & 21h48m19.5s & -34d57m05s & 0.016151 & 16.36 & 26.19 \\
NGC\,7469 & 23h03m15.6s & +08d52m26s & 0.016317 & 26.49 & 36.15 \\
Mrk\,331  & 23h51m26.8s & +20d35m10s & 0.018483 & 18.30 & 23.44 \\
\enddata
\tablecomments{Given by NASA Extragalactic Database (NED).}
\end{deluxetable}

\begin{deluxetable}{ll}
\footnotesize
\rotate
\tablecaption{Other far-IR/submm/mm photometric data available for the local LIRG sample\label{local-photometry}}
\tablewidth{0pt}
\tablehead{\colhead{Source Name} & \colhead{far-IR/submm/mm Photometric Data}\\
\colhead{} &  \colhead{Jy}}
\startdata
NGC\,520  & $\rm S_{850um} = 0.325 \pm 0.050,~ \rm S^{(a)}_{1.3mm}= 0.050 \pm 0.025 $\\
NGC\,1614 & $\rm S_{850um} = 0.219 \pm 0.032,~ \rm S^{(a)}_{1.3mm}= 0.011 \pm 0.003 $\\
NGC\,2339 & $\rm S_{450um} = 0.849 \pm 0.078,~ \rm S_{800um}= 0.079 \pm 0.015,~\rm S_{1.3mm}= 0.019 \pm 0.002 $\\
NGC\,2388 & \nodata \\
NGC\,2623 & $\rm S_{850um} = 0.091 \pm 0.014 $\\
UGC\,5101 & $\rm S_{450um} = 1.43 \pm 0.304,~ S_{800um} = 0.143 \pm 0.025,~ S_{1.1mm} = 0.068 \pm 0.012$\\
NGC\,4102 & $\rm S^{(b)}_{450um} = 1.79 \pm 0.450,~ S^{(b)}_{800um} = 0.212 \pm 0.028,~ S^{(b)}_{1.1mm}= 0.055 \pm 0.007,~ S^{(b)}_{1.3mm}= 0.050 \pm 0.006$\\
NGC\,4194 & \nodata \\
NGC\,4418 & $\rm S^{(b)}_{450um}= 1.34 \pm 0.350,~ S^{(b)}_{800um} = 0.240 \pm 0.020,~ S_{850um} = 0.255 \pm 0.037,~ S^{(b)}_{1.1mm} = 0.085 \pm 0.009$\\
          & $\rm S^{(b)}_{1.3mm} = 0.072 \pm 0.012$\\
Mrk\,231  & $\rm S_{120um}= 24.32 \pm 30\%,~ S_{150um} = 14.74 \pm 30\%,~ S_{180um} = 9.75 \pm 30\%,~ S_{200um} = 6.88 \pm 30\%$\\
          & $\rm S_{450um}= 0.513 \pm 0.134,~ S^{(b)}_{800um} = 0.085 \pm 0.012,~ S^{(b)}_{1.1mm} = 0.045 \pm 0.012,~ S_{1.25mm} =0.029 \pm 0.008$\\
          & $\rm S^{(b)}_{1.3mm}= 0.038 \pm 0.008$\\
NGC\,4818 & \nodata  \\
NGC\,5135 & \nodata  \\
Mrk\,273  & $\rm S_{120um}= 19.99 \pm 30\%,~ S_{150um} = 13.10 \pm 30\%,~ S_{180um} = 8.69 \pm 30\%,~ S_{200um} = 7.40 \pm 30\%$\\
          & $\rm S_{450um} = 0.707 \pm 0.200,~ S_{800um} = 0.084 \pm 0.022,~ S_{1.1mm} = 0.051 \pm 0.013,~ S^{(a)}_{1.3mm} = 0.020 \pm 0.002 $\\
NGC\,6000 & $\rm S^{(b)}_{450um} = 1.460 \pm 0.650,~ S^{(b)}_{800um} = 0.115 \pm 0.027,~ S^{(b)}_{1.1mm} = 0.055 \pm 0.008,~ S^{(b)}_{1.3mm} = 0.018 \pm 0.008$\\
NGC\,6240 & $\rm S_{120um} = 25.90 \pm 30\%,~ S_{150um} = 18.91 \pm 30\%,~ S_{180um} = 12.73 \pm 30\%,~ S_{200um} = 9.00 \pm 30\%$\\
          & $\rm S_{450um} = 1.00 \pm 30\%,~ S_{850um} = 0.150 \pm 30\%,~ S^{(a)}_{1.3mm} = 0.021 \pm 0.010 $\\
IC\,5135  & $\rm S_{1.3mm} = 0.033 \pm 0.004 $\\
NGC\,7469 & $\rm S_{850um} = 0.192 \pm 0.027 $\\
Mrk\,331  & $\rm S_{850um} = 0.132 \pm 0.025 $\\
\enddata
\tablecomments{Given by NED unless noted otherwise.}
\tablerefs{(a)\ Benford\ 1999; (b)\ Roche \& Chandler\ 1993.}
\end{deluxetable}

\begin{deluxetable}{lcccccccc}
\footnotesize
\tablecaption{350\um flux and derived properties for the local LIRG sample\label{local-s350-sed}}
\tablewidth{0pt}
\tablehead{\colhead{Source Name} & \colhead{$\rm S_{350}$} & \colhead{$\rm \sigma_{350}$} &
\colhead{\Td} & \colhead{$\rm \sigma(T_d)$}  & \colhead{\bet} & \colhead{$\rm \sigma(\beta)$} & \colhead{log\,\Lfir} & \colhead{log\,\Md}\\
\colhead{} & \colhead{Jy}  & \colhead{Jy} & \colhead{\K} & \colhead{\K} & \colhead{} & \colhead{} & \colhead{\Lsun} & \colhead{\Msun}}
\startdata
NGC\,520   &    3.01 &  0.02  &  37.8  & 4.3  & 1.5  & 0.2 & 10.76 &  6.94\\
NGC\,1614  &    1.18 &  0.05  &  40.7  & 5.0  & 1.9  & 0.2 & 11.37 &  7.30\\
NGC\,2339  &    1.78 &  0.04  &  32.1  & 3.2  & 2.1  & 0.2 & 10.54 &  7.01\\
NGC\,2388  &    1.87 &  0.10  &  37.9  & 3.2  & 1.5  & -   & 11.03 &  7.21\\
NGC\,2623  &    1.43 &  0.05  &  41.8  & 6.6  & 1.6  & 0.3 & 11.38 &  7.32\\
UGC\,5101  &    2.20 &  0.10  &  34.1  & 3.6  & 1.6  & 0.2 & 11.86 &  8.27\\
NGC\,4102  &    4.38 &  0.04  &  32.1  & 3.0  & 2.1  & 0.2 & 10.10 &  6.56\\
NGC\,4194  &    1.19 &  0.03  &  45.3  & 4.7  & 1.5  & -   & 10.68 &  6.47\\
NGC\,4418  &    2.03 &  0.07  &  62.9  & 12.4 & 0.9  & 0.2 & 10.81 &  6.14\\
Mrk\,231   &    1.73 &  0.07  &  51.3  & 5.6  & 1.4  & 0.1 & 12.24 &  7.79\\
NGC\,4818  &    2.73 &  0.05  &  36.0  & 3.2  & 1.5  & -   &  9.91 &  6.20\\
NGC\,5135  &    3.80 &  0.07  &  33.0  & 2.6  & 1.5  & -   & 11.10 &  7.59\\
Mrk\,273   &    1.34 &  0.08  &  42.7  & 4.0  & 1.6  & 0.1 & 11.98 &  7.88\\
NGC\,6000  &    4.26 &  0.20  &  29.2  & 2.6  & 2.4  & 0.2 & 10.81 &  7.48\\
NGC\,6240  &    1.10 &  0.05  &  39.4  & 3.4  & 1.7  & 0.1 & 11.64 &  7.69\\
IC\,5135   &    2.16 &  0.10  &  35.5  & 3.9  & 1.6  & 0.2 & 11.17 &  7.49\\
NGC\,7469  &    2.23 &  0.20  &  42.3  & 6.9  & 1.3  & 0.3 & 11.35 &  7.31\\
Mrk\,331   &    1.69 &  0.04  &  41.5  & 6.5  & 1.4  & 0.3 & 11.29 &  7.30\\
\enddata
\tablecomments{\bet=1.5 is assumed in the SED fitting for NGC\,2388, NGC\,4194, NGC\,4818 and NGC\,5135. 
The luminosity distance used in the calculation of the far-IR luminosity is derived 
from the redshift, by assuming a flat cosmology with $\rm H_0=70\,{km\,s^{-1}\,Mpc^{-1}}$, 
$\rm \Omega_m=0.30$ and $\rm \Omega_{\Lambda}=0.70$.}
\end{deluxetable}

\begin{figure}
\begin{center}
\begin{tabular}{ccc}
{\includegraphics[width=2.5in, angle=90]{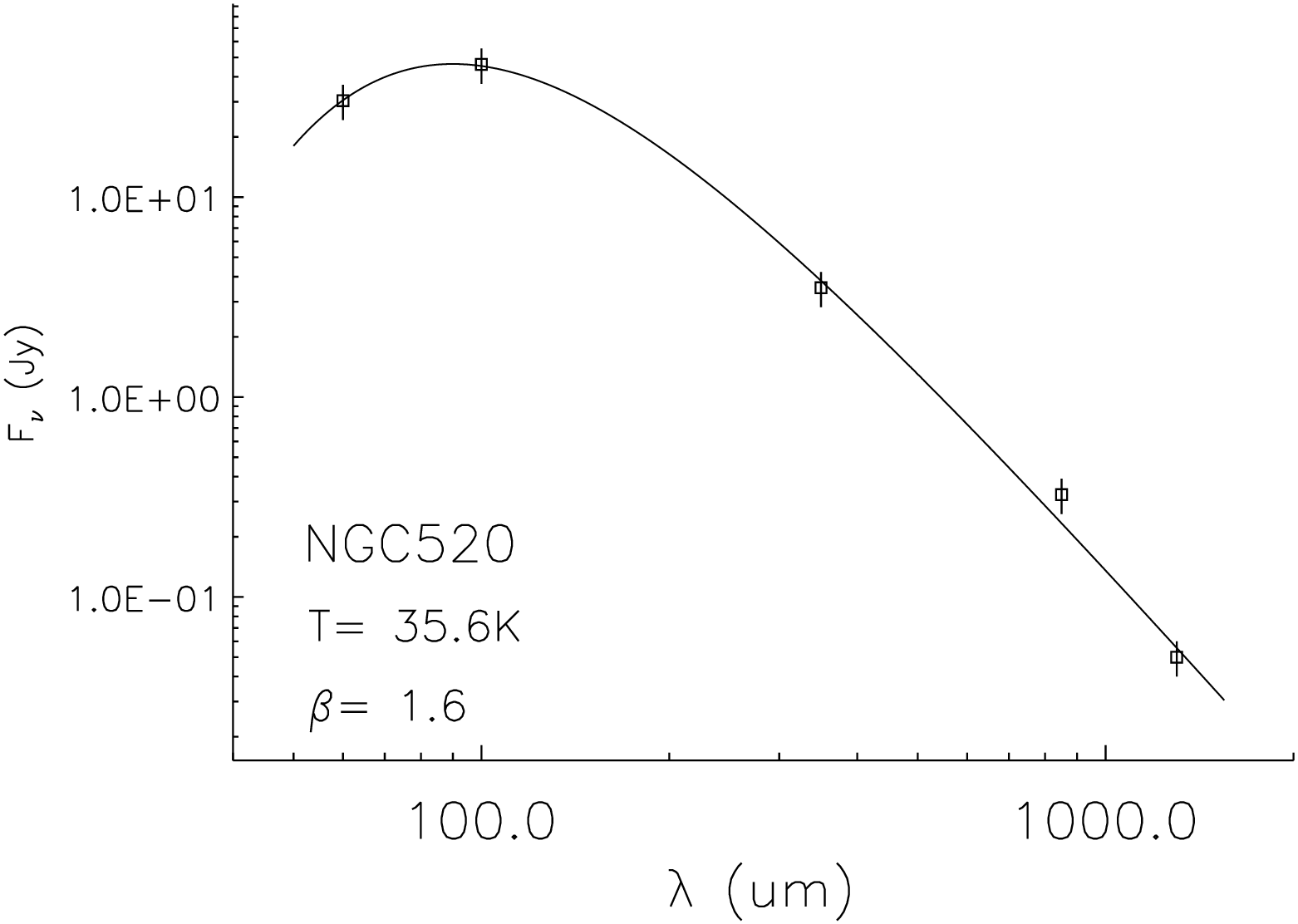}}&
{\includegraphics[width=2.5in, angle=90]{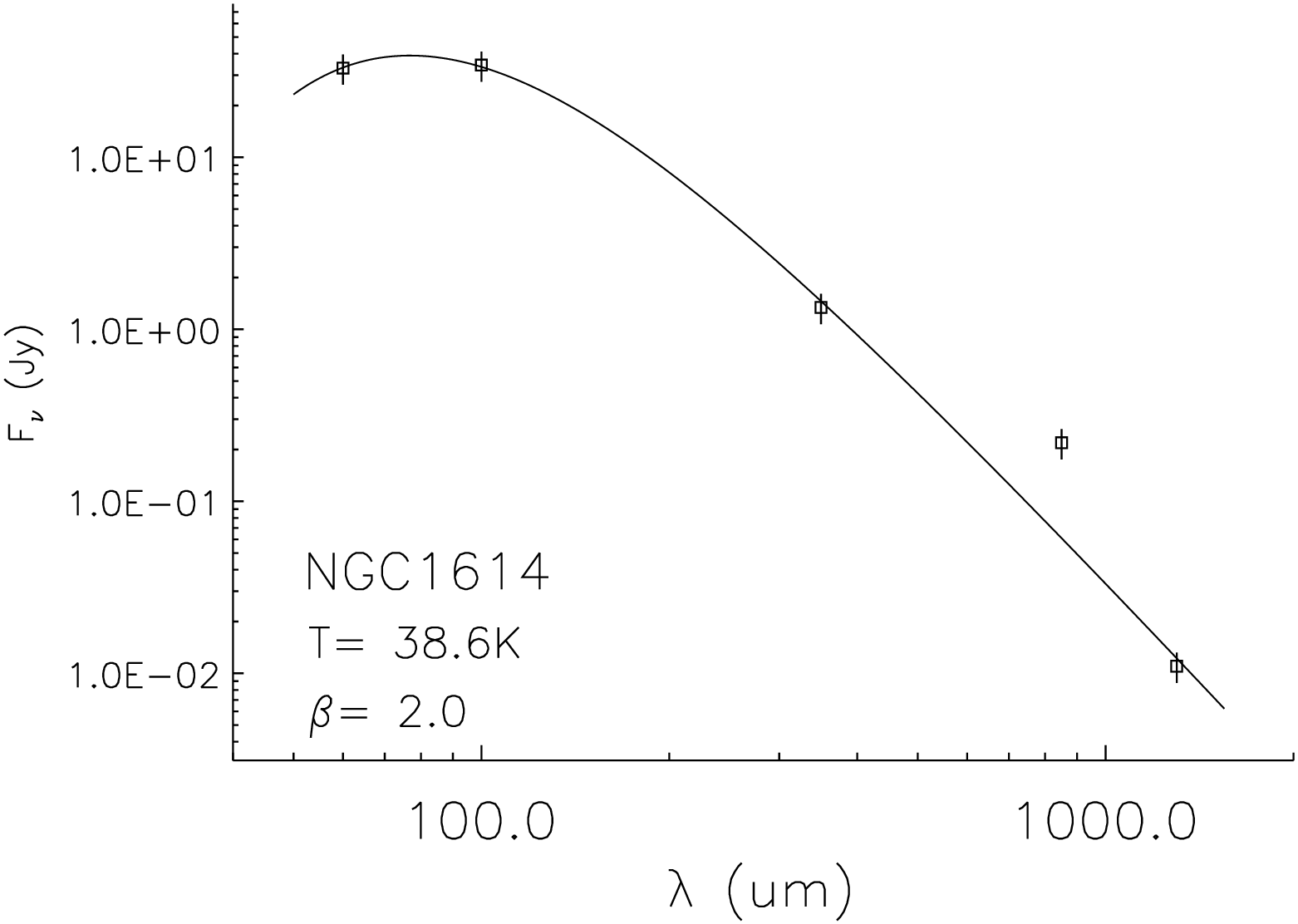}}\\
{\includegraphics[width=2.5in, angle=90]{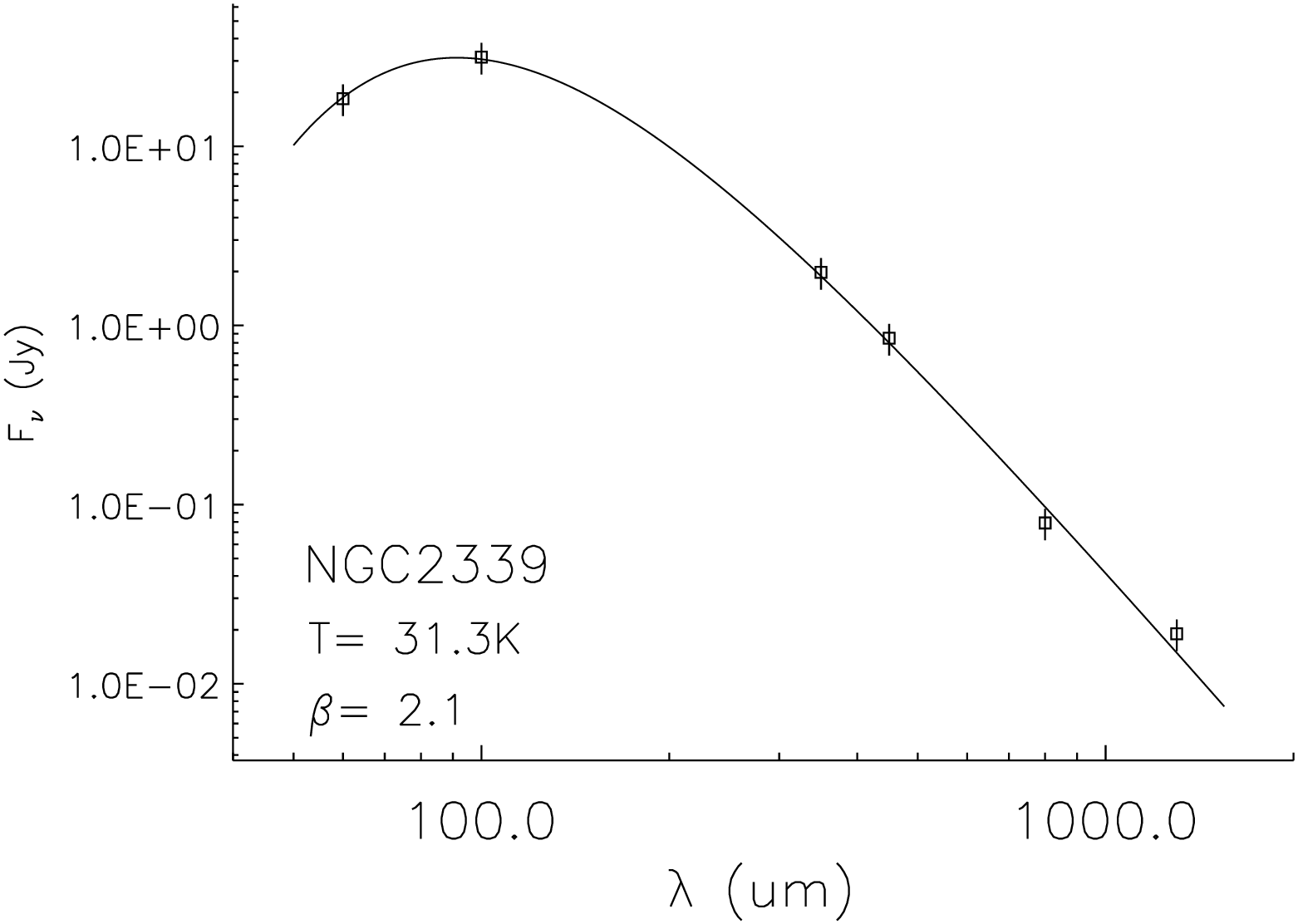}}&
{\includegraphics[width=2.5in, angle=90]{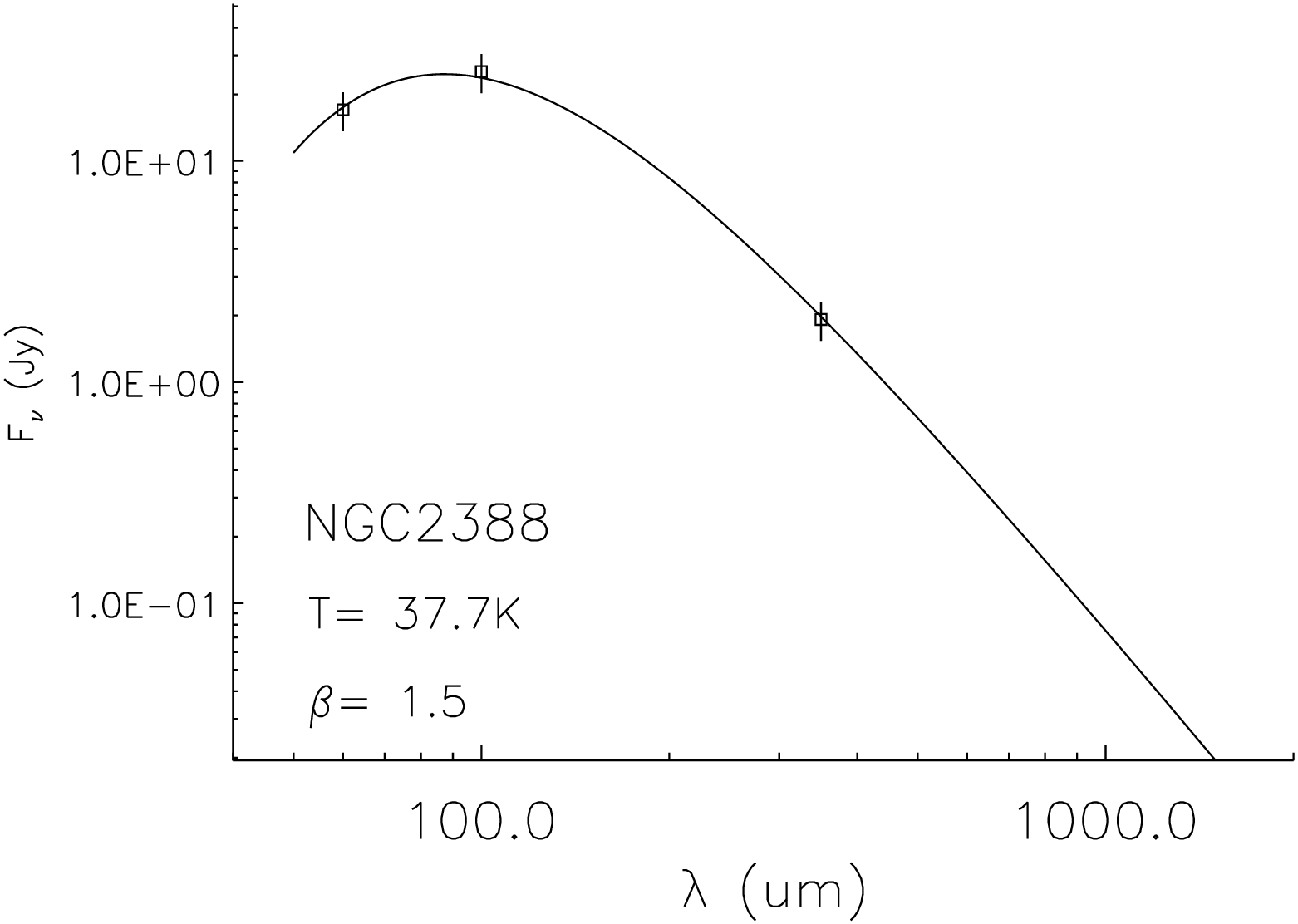}}\\
{\includegraphics[width=2.5in, angle=90]{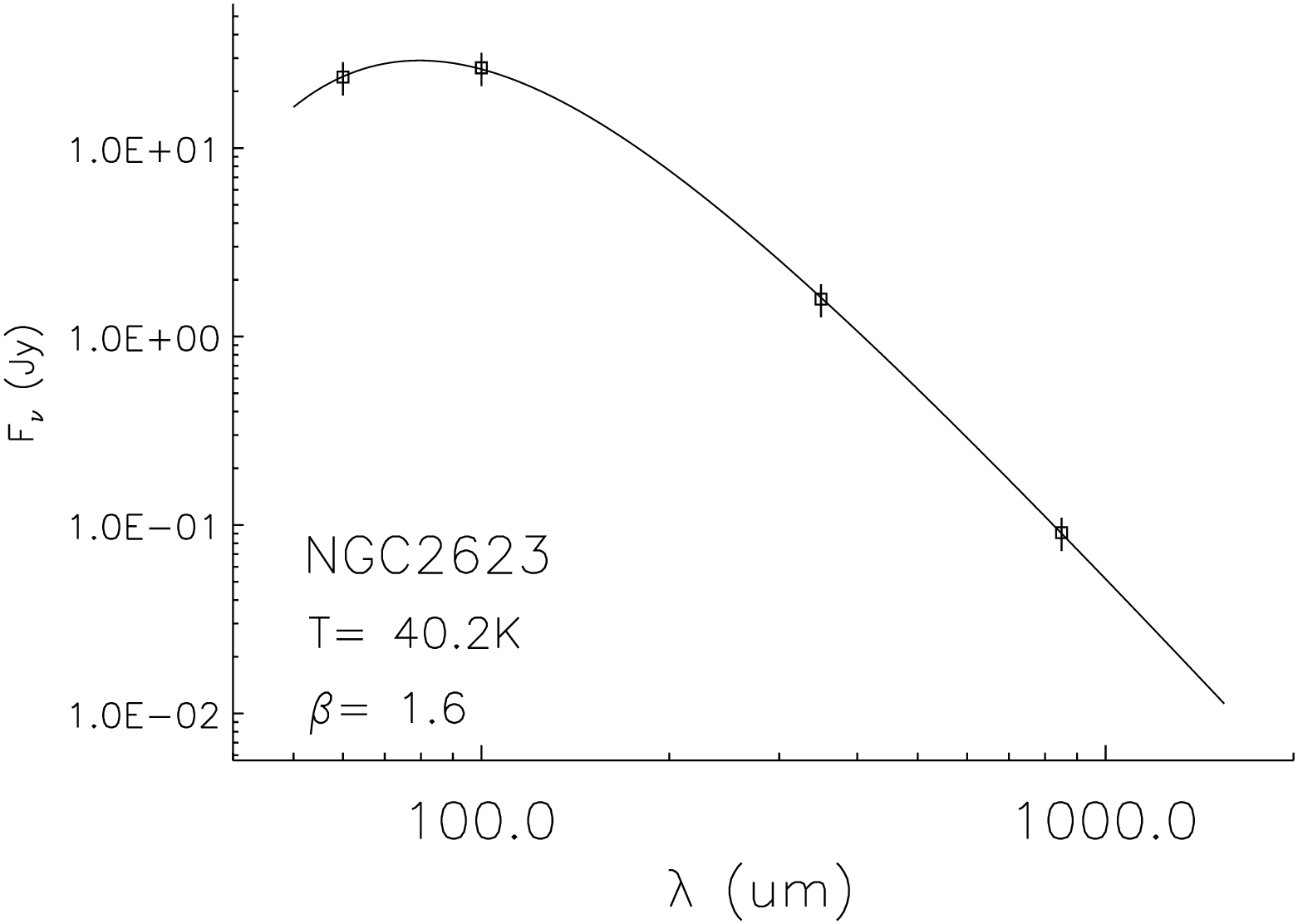}}&
{\includegraphics[width=2.5in, angle=90]{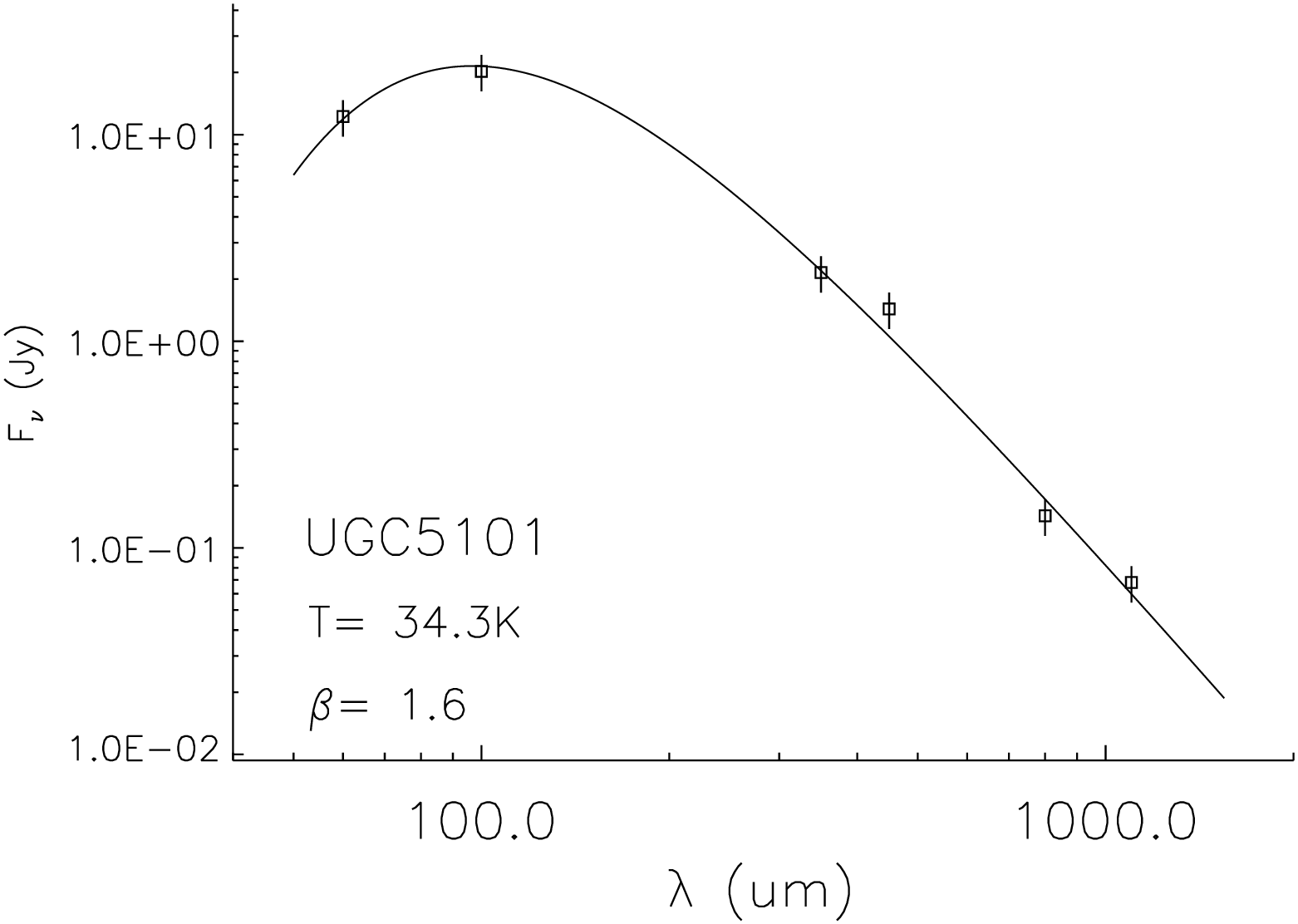}}\\
\end{tabular}
\caption{SED fits for galaxies in the local LIRG sample.}\label{local-sedplot}
\end{center}
\end{figure}

\begin{figure*}
\setcounter{figure}{1}
\begin{center}
\begin{tabular}{ccc}
{\includegraphics[width=2.5in, angle=90]{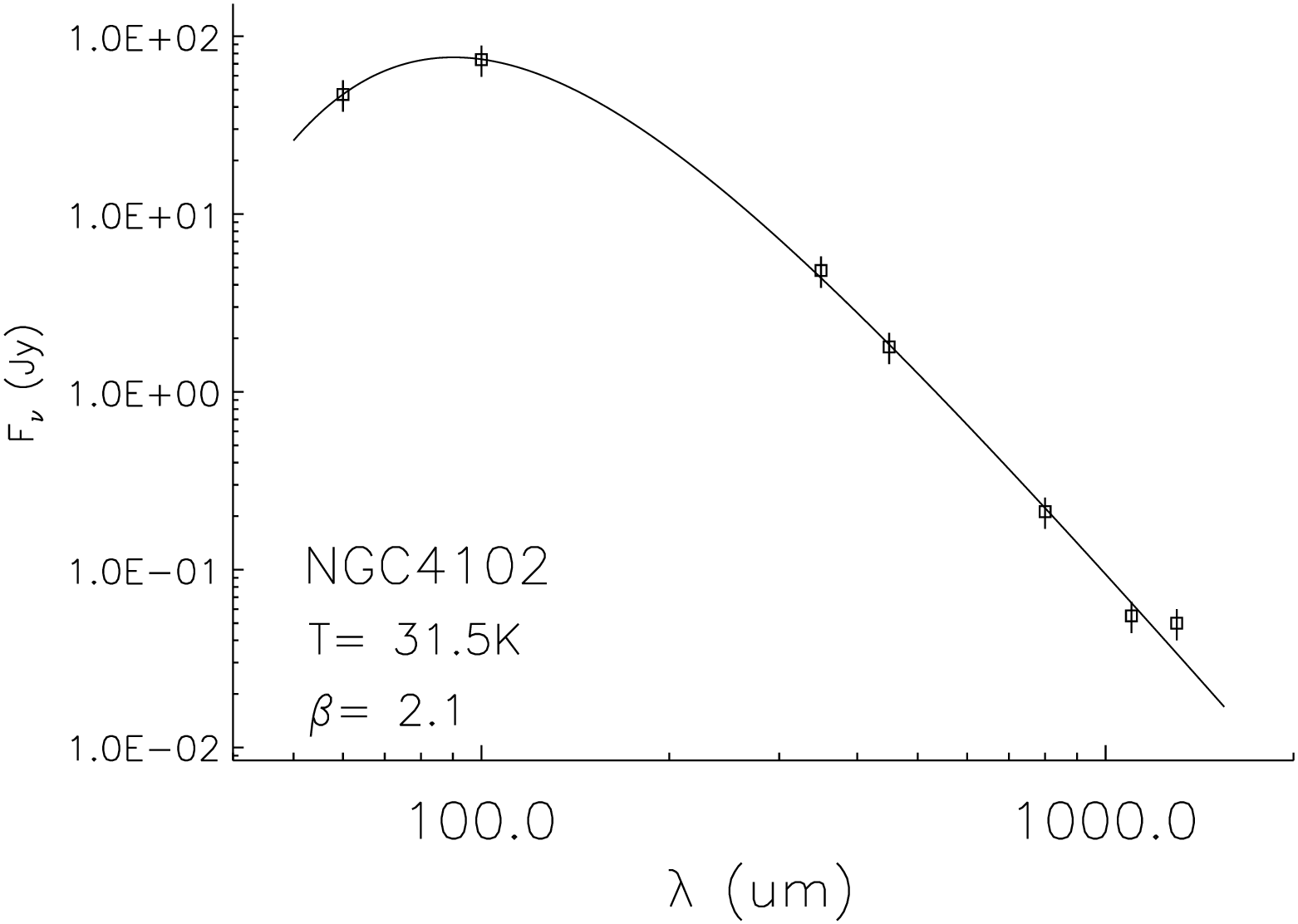}}&
{\includegraphics[width=2.5in, angle=90]{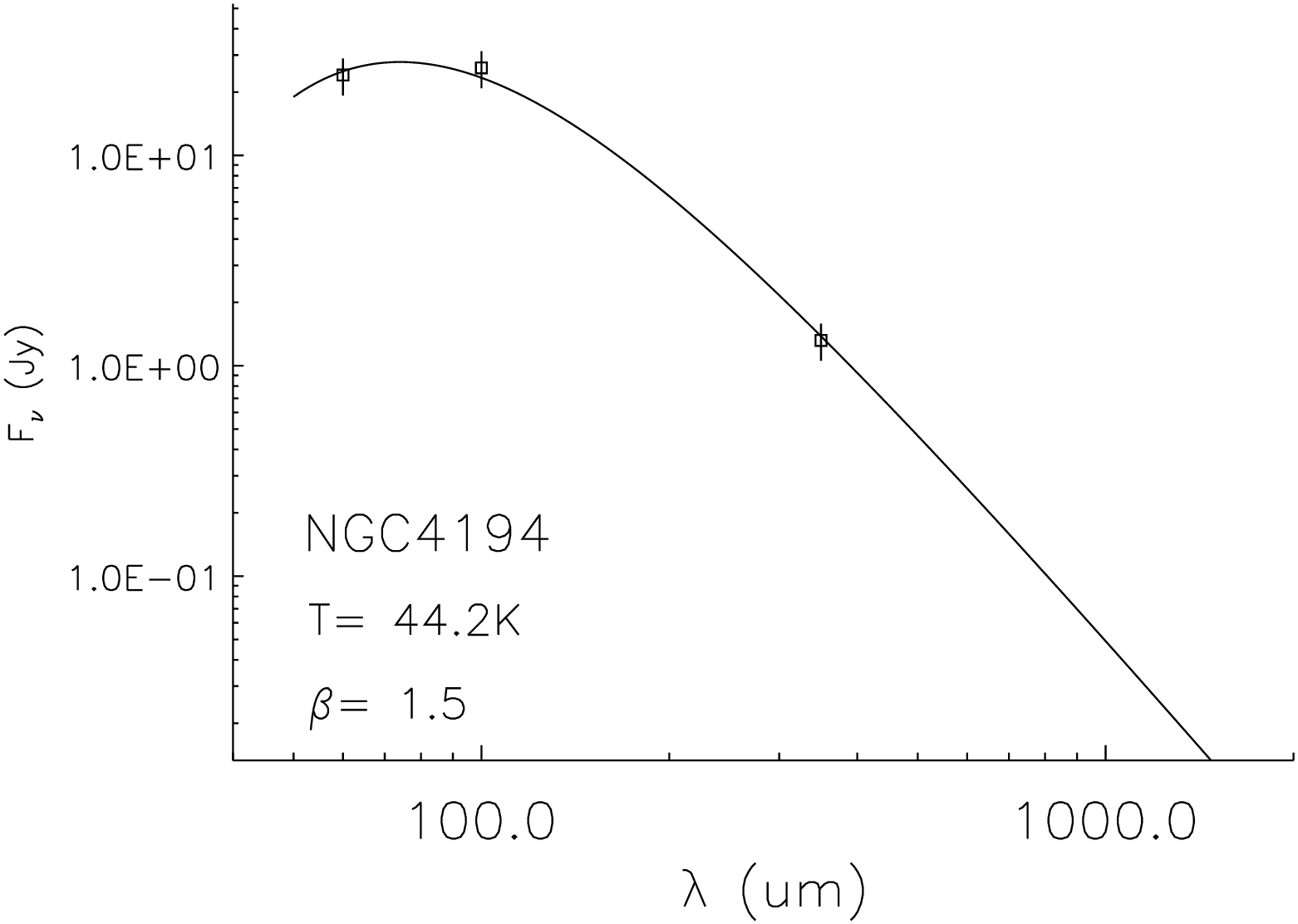}}\\
{\includegraphics[width=2.5in, angle=90]{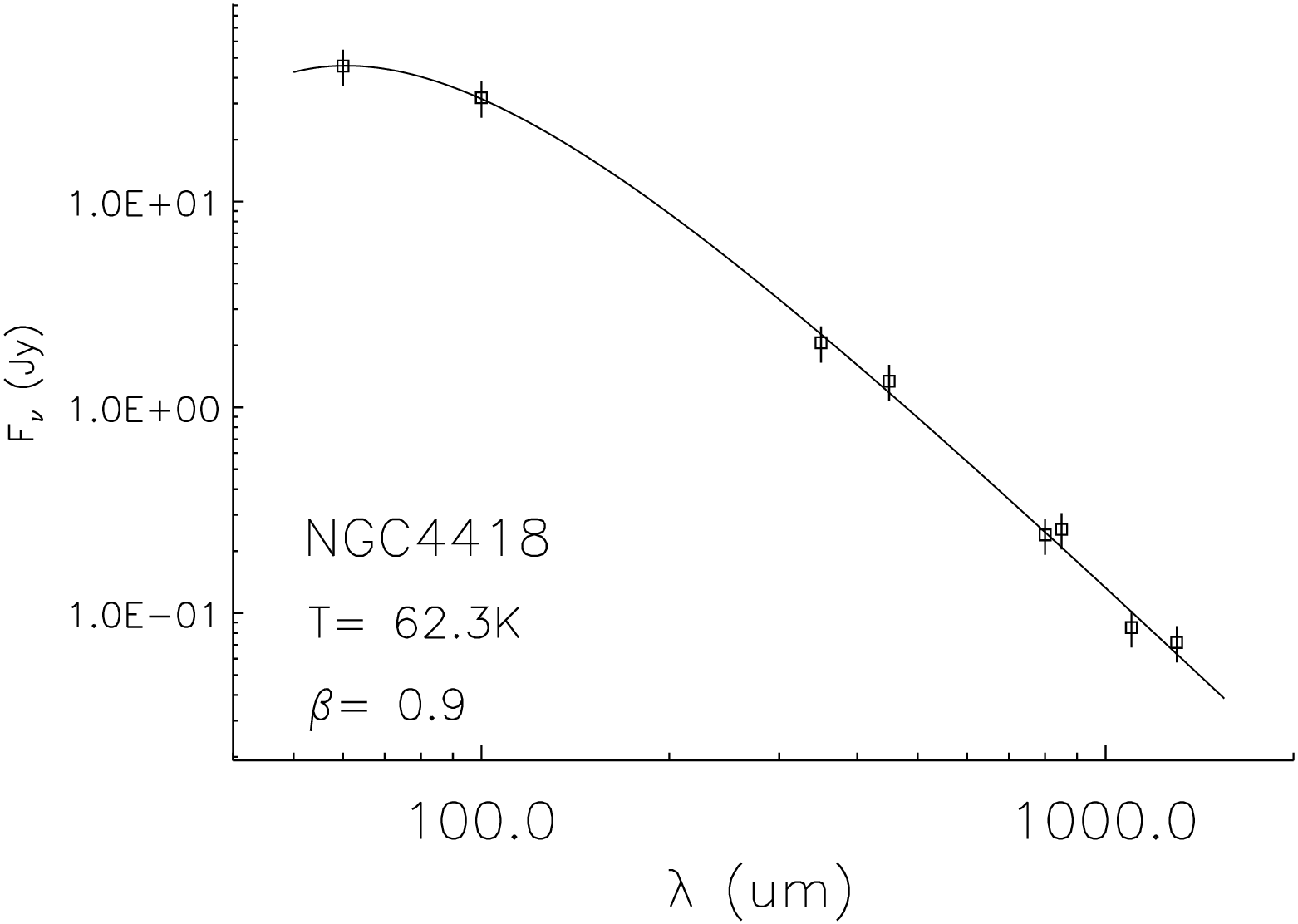}}&
{\includegraphics[width=2.5in, angle=90]{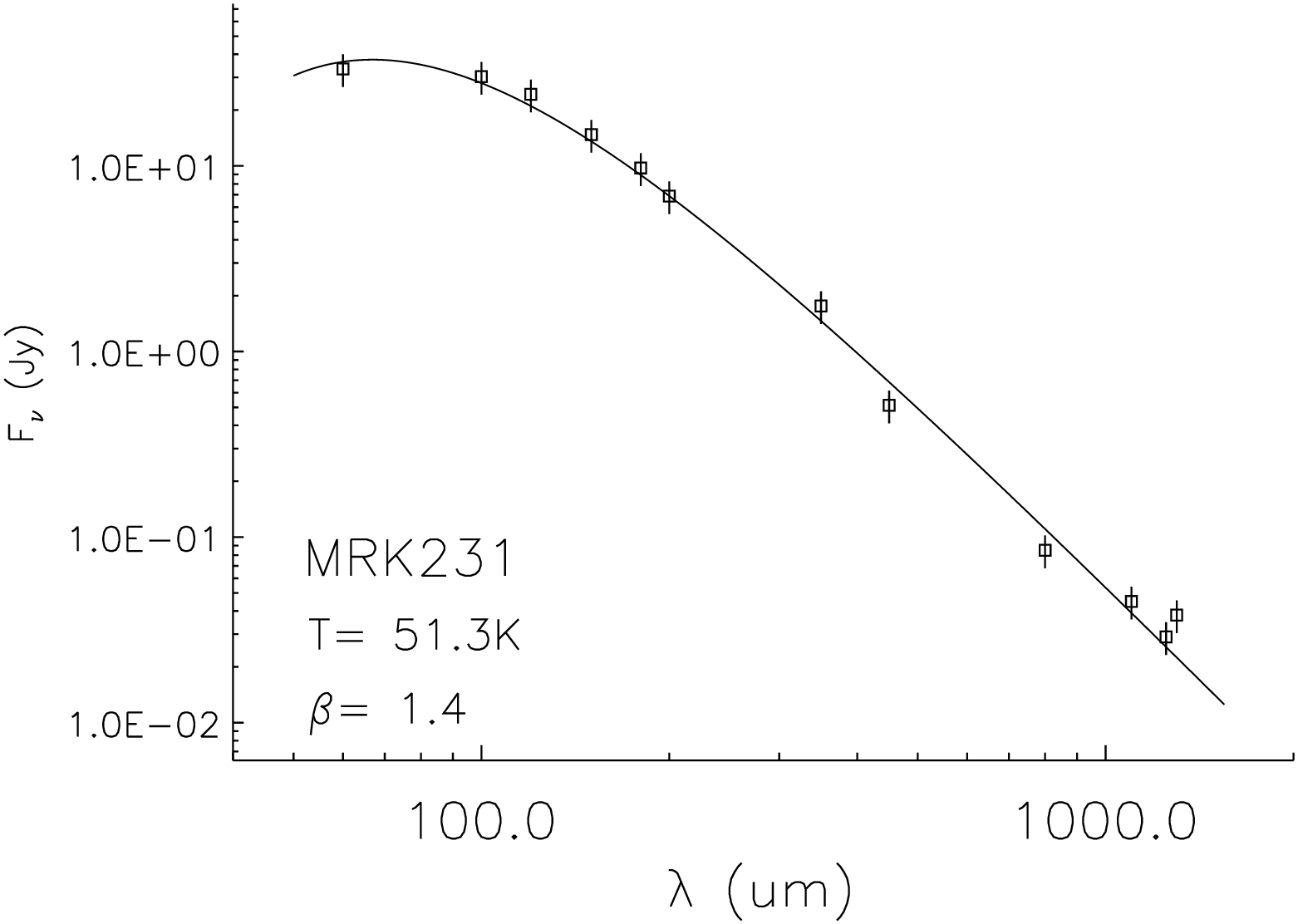}}\\
{\includegraphics[width=2.5in, angle=90]{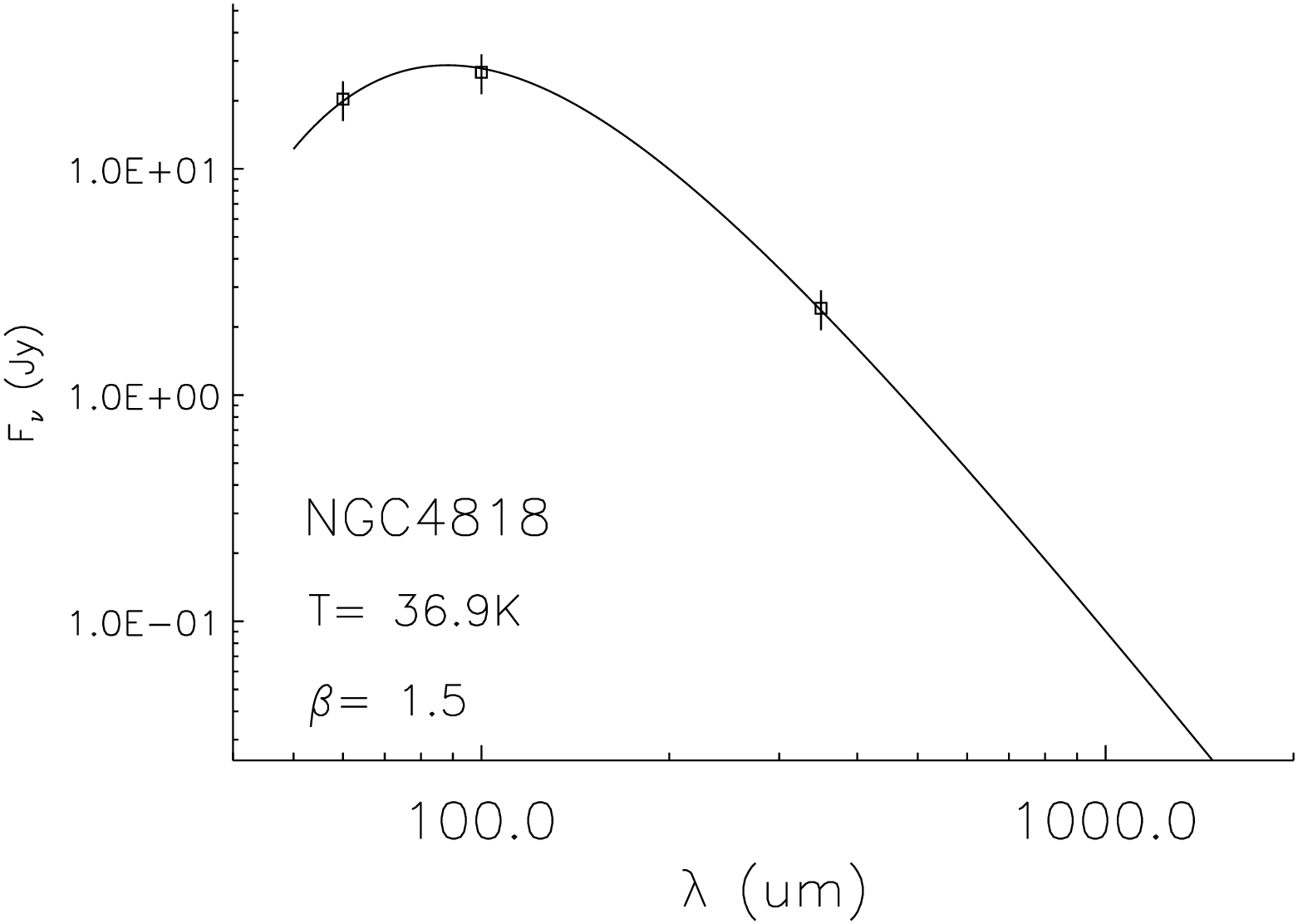}}&
{\includegraphics[width=2.5in, angle=90]{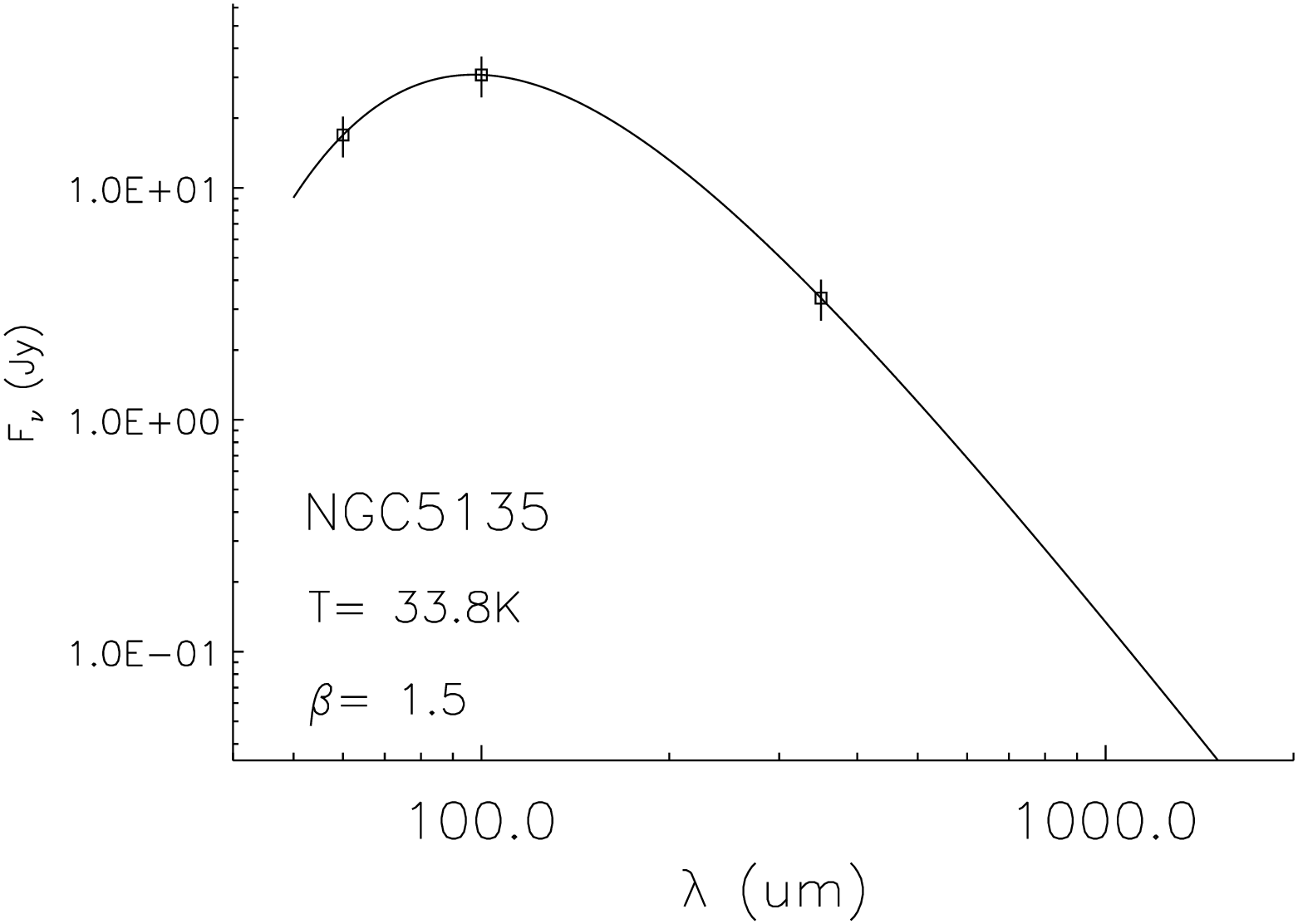}}\\
\end{tabular}
\caption{Continued}
\end{center}
\end{figure*}

\begin{figure*}
\setcounter{figure}{1}
\begin{center}
\begin{tabular}{ccc}
{\includegraphics[width=2.5in, angle=90]{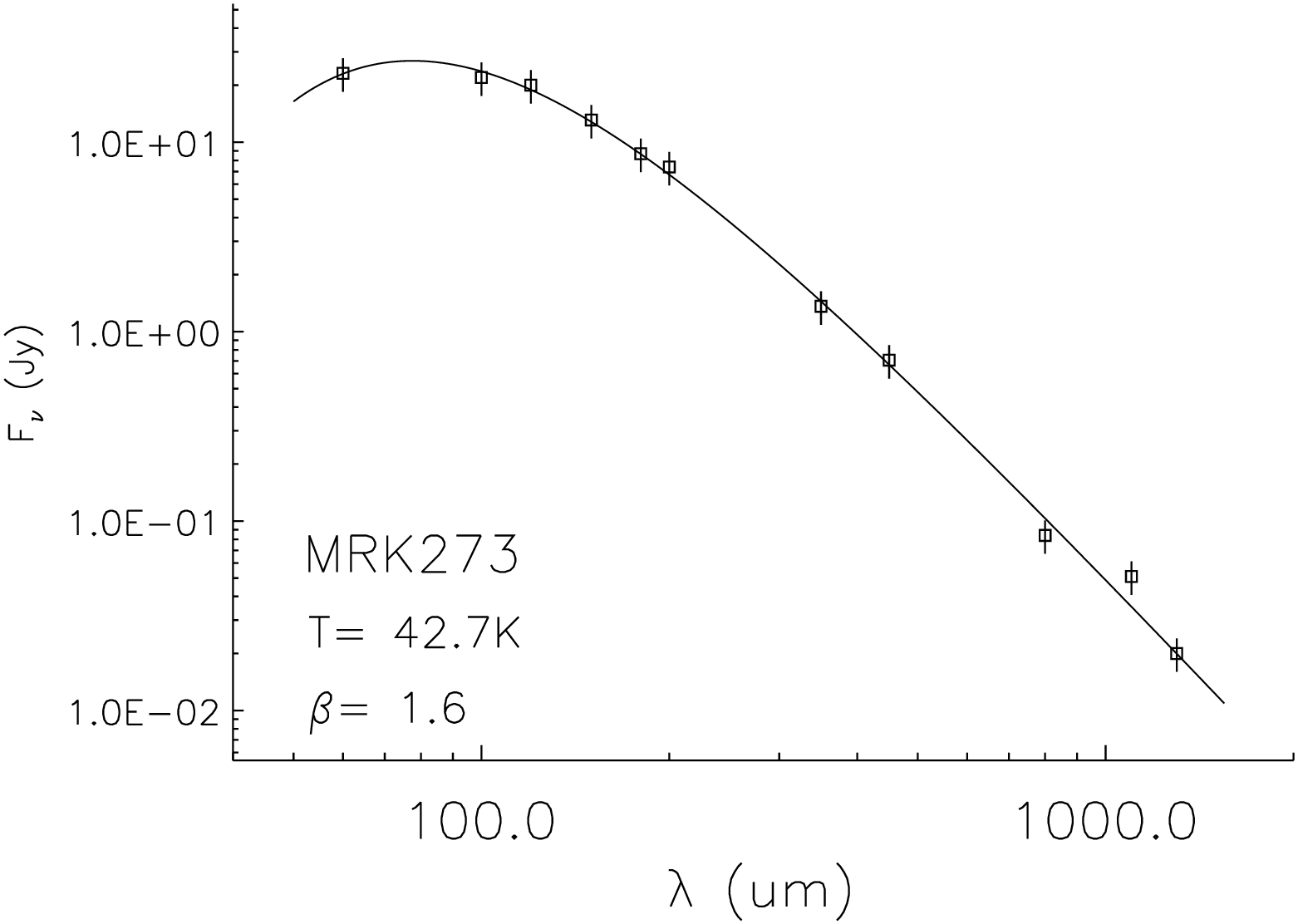}}&
{\includegraphics[width=2.5in, angle=90]{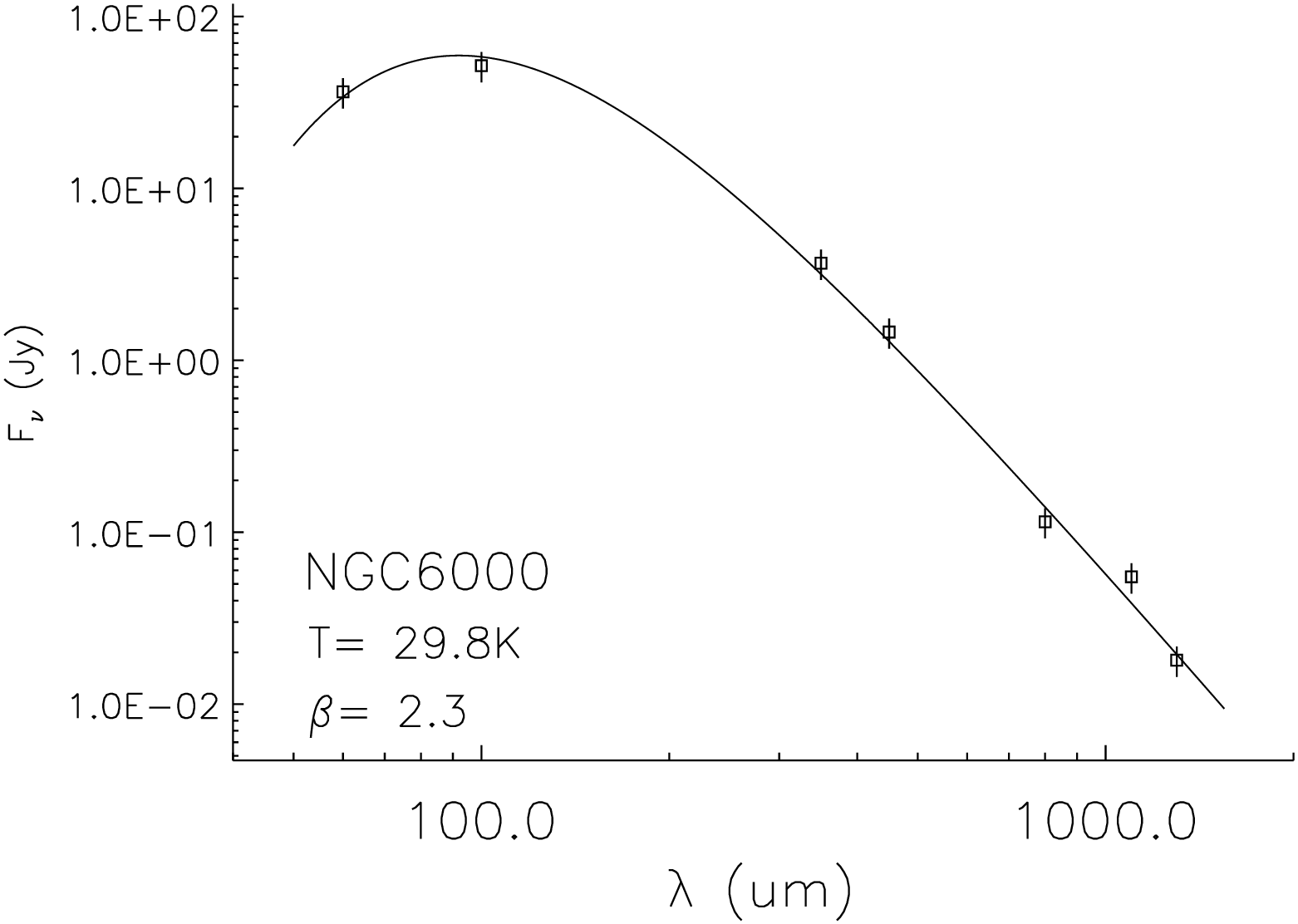}}\\
{\includegraphics[width=2.5in, angle=90]{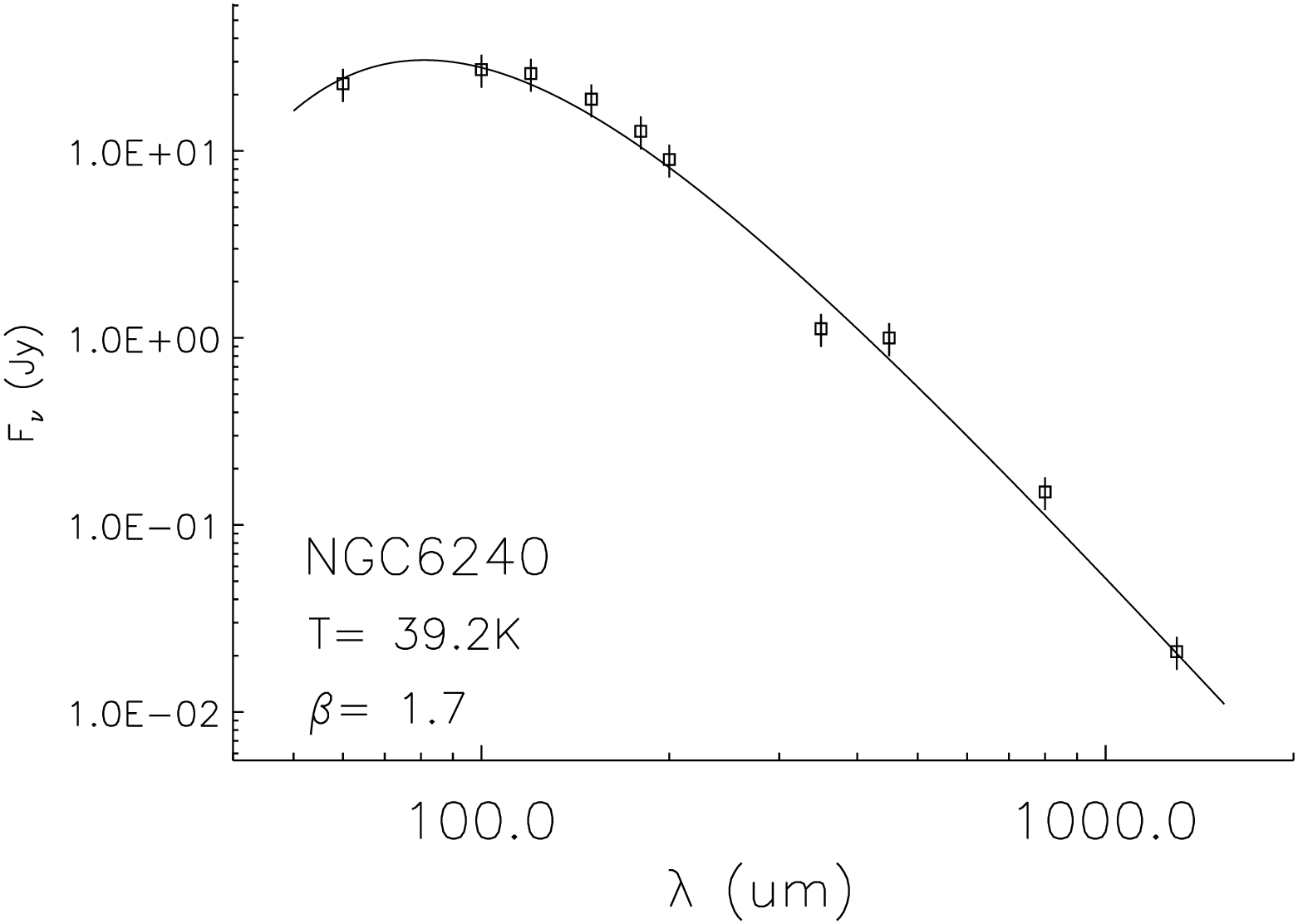}}&
{\includegraphics[width=2.5in, angle=90]{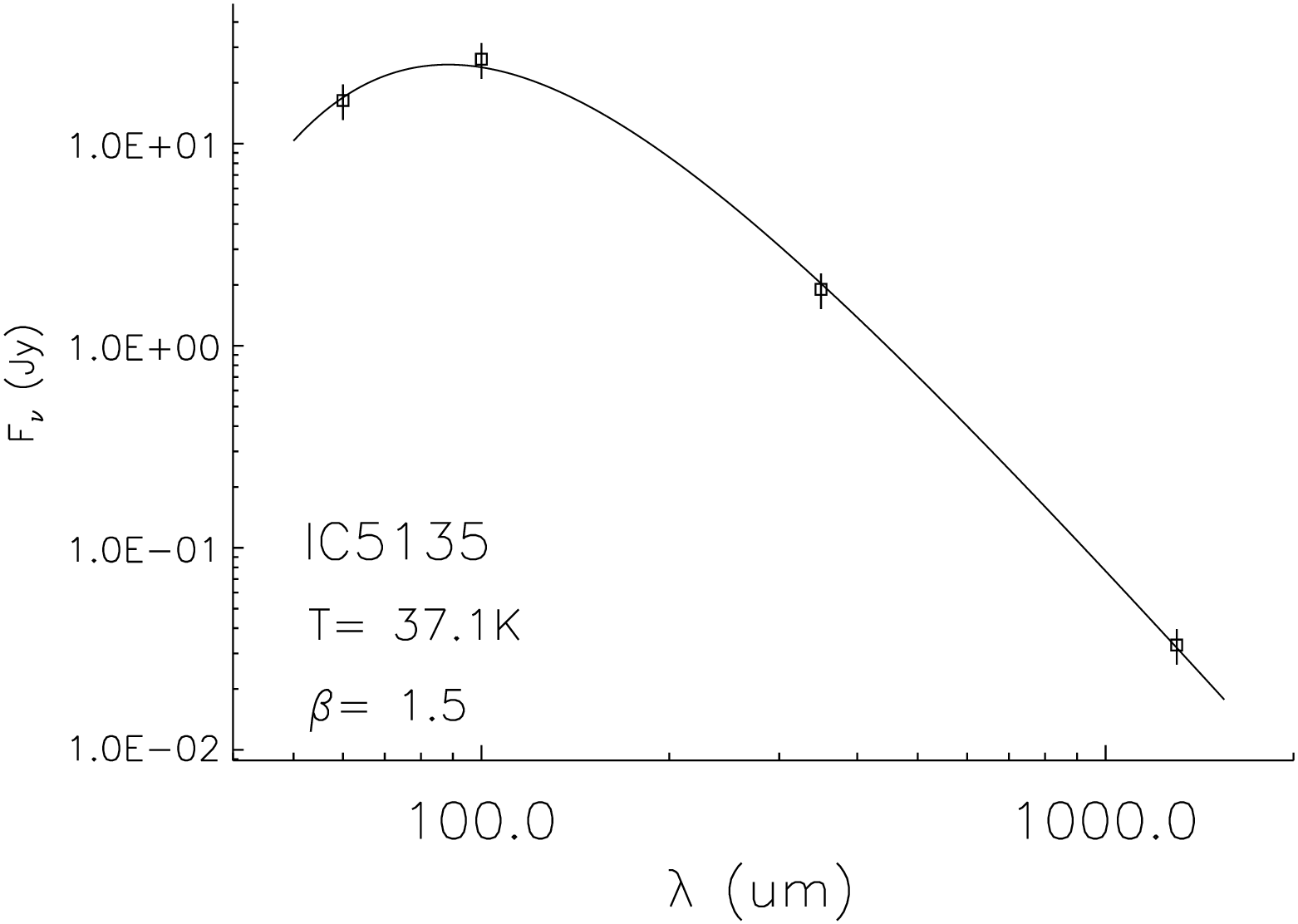}}\\
{\includegraphics[width=2.5in, angle=90]{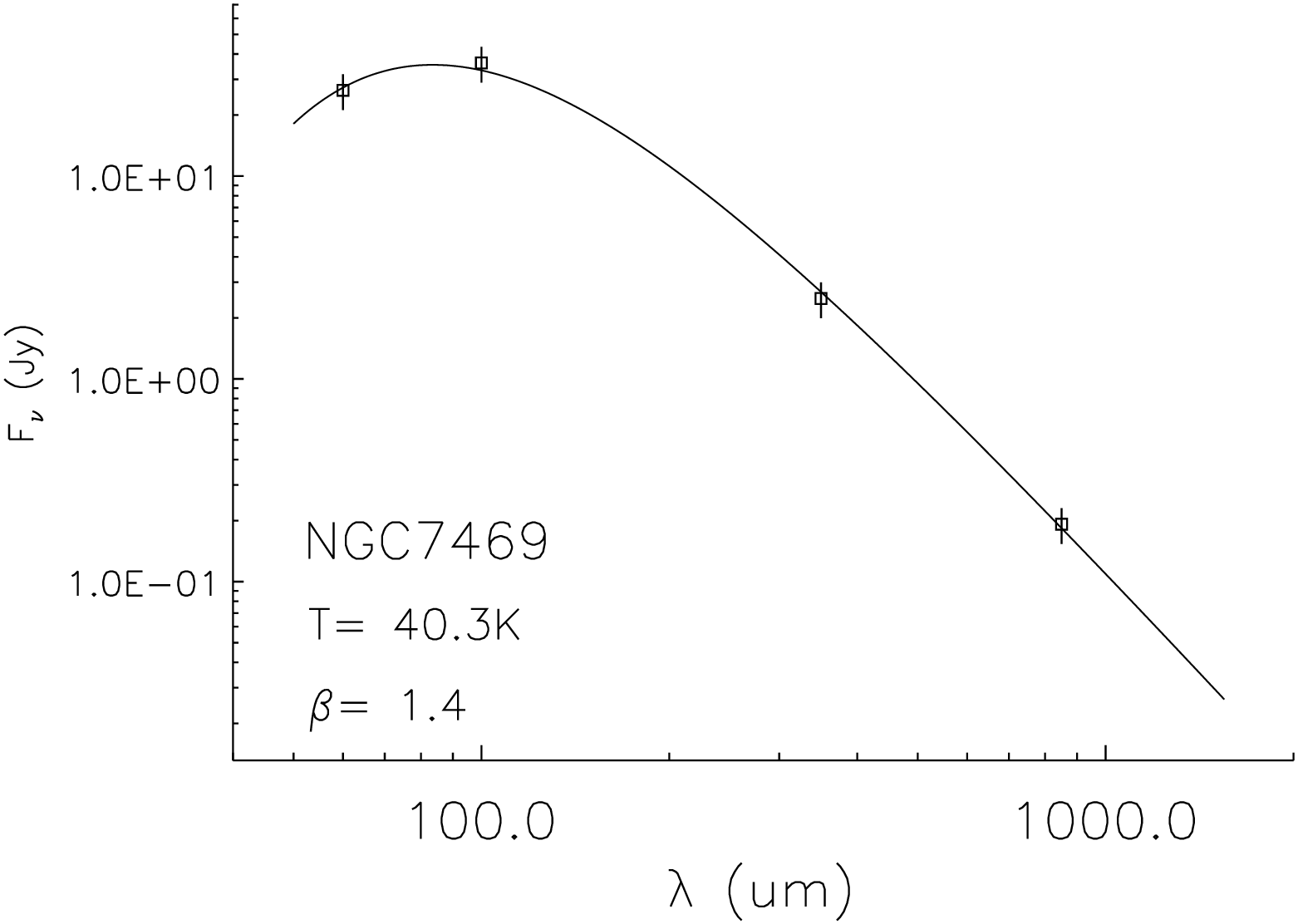}}&
{\includegraphics[width=2.5in, angle=90]{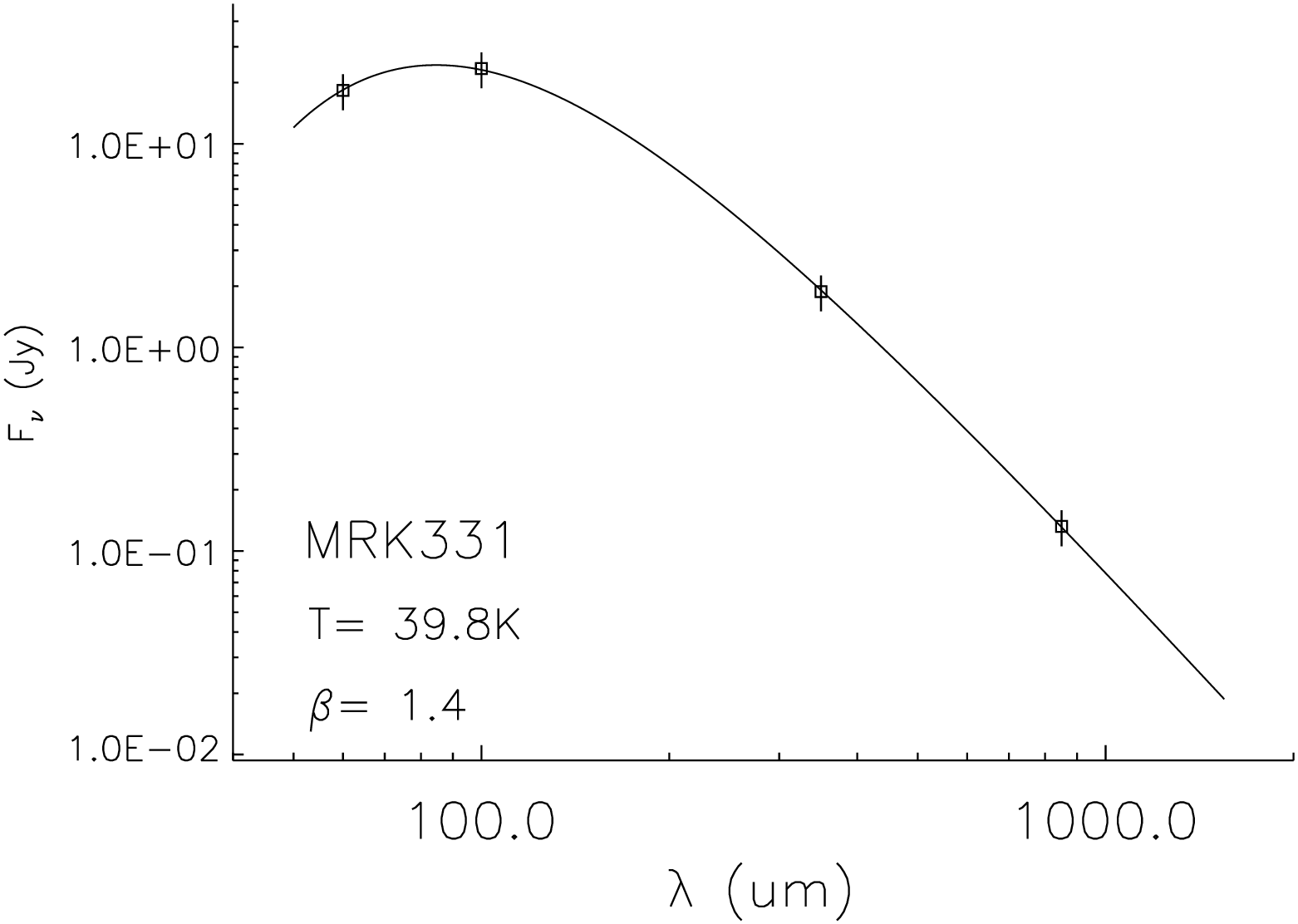}}\\
\end{tabular}
\caption{Continued}
\end{center}
\end{figure*}

\end{document}